\documentclass[10pt,journal,letterpaper]{IEEEtran}
\usepackage[utf8]{inputenc}
\usepackage[T1]{fontenc}
\usepackage{graphicx}
\usepackage{epsfig}
\usepackage[bottom]{footmisc}
\usepackage{cite}
\usepackage{amsmath, amssymb, amsfonts}
\usepackage{fancyhdr}
\usepackage{float}
\usepackage{algorithm}
\usepackage{algpseudocode}
\usepackage{algorithmicx}
\usepackage{textcomp} 
\usepackage{multirow}
\usepackage[xcdraw, table]{xcolor}
\usepackage{booktabs}
\usepackage{caption}
\usepackage{subcaption}
\usepackage{rotating}
\usepackage{tablefootnote}  
\usepackage{tikz}
\usepackage{mathtools}
\usepackage{multirow}
\usepackage{amssymb} 

\def\BibTeX{{\rm B\kern-.05em{\sc i\kern-.025em b}\kern-.08em
    T\kern-.1667em\lower.7ex\hbox{E}\kern-.125emX}}

\begin{document}

% \title{Reliable Virtual PIN Entry on Passive NFC Cards via Domain Adaptation and Uncertainty Quantification}
\title{Adaptive Virtual PIN-Button Recognition on Passive NFC Cards via Domain Adaptation and Uncertainty Quantification}

\author{Dickson Akuoko Sarpong,~\IEEEmembership{Graduate Student Member,~IEEE},
and Hongzhi Guo,~\IEEEmembership{Senior Member,~IEEE}%
\thanks{This article extends work presented in part at the
2026 IEEE International Conference on Sensing,
Communication, and Networking (SECON).}%
\thanks{This work was supported in part by the National
Science Foundation under Grant CNS-2341846 and Grant CNS-2310856.}%
\thanks{The authors are with the
School of Computing, University of Nebraska--Lincoln,
Lincoln, NE 68588 USA (e-mail:
dakuokosarpong2@huskers.unl.edu; hguo10@unl.edu).}%
\thanks{Corresponding author: Dickson Akuoko Sarpong.}%
}

\maketitle

\begin{abstract}
Passive Near Field Communication (NFC) cards are widely used for access control and authentication but provide no native interface for reliable user input. Radio-frequency (RF)-based interaction recognition offers a way to enable input on these cards, but hardware-dependent signal variations across cards and readers limit recognition reliability. This paper presents a practical virtual input framework that enables reliable PIN-button entry on passive NFC cards using a 12-button interface without modifying the commercial NFC card or the ISO/IEC 15693 communication protocol. We develop a multi-domain adaptation framework that jointly learns interaction representations invariant to both card and reader variations. To improve decision reliability, we introduce a model-agnostic uncertainty quantification framework based on conformal prediction that estimates recognition confidence. When confidence is insufficient, the system requests re-input; otherwise, accepted inputs are immediately confirmed through reader feedback, enabling reliable user interaction. A Software-Defined Radio (SDR)-based ISO/IEC 15693 testbed is developed using multiple commercial NFC cards and reader configurations. Experimental results show that domain adaptation improves recognition accuracy from 93.75\% to 95.42\% for the deep learning (DL) model and from 85.00\% to 89.11\% for the Random Forest model. Conformal prediction further enables adaptive high-confidence recognition, achieving average recognition accuracies of 96.85\% and 90.55\% for the DL and Random Forest models, respectively, across diverse hardware configurations.
\end{abstract}


% \printnomenclature

\section{Introduction}
\label{sec:introduction}

Secure access control is essential for protecting restricted facilities and sensitive resources. Many access-control systems rely on identity verification to ensure that only authorized users gain access. For example, some systems require users to present an identification (ID) card and enter a Personal Identification Number (PIN) on a separate keypad~\cite{2022PersonalIV}. Integrating PIN input directly into the identification card could eliminate the need for a separate keypad. Passive radio-frequency (RF) interaction provides a potential approach to such integration by enabling battery-free touch, gesture, and numerical input through changes in impedance, resonance, backscatter, magnetic coupling, and related RF responses~\cite{10.1145/3117811.3117818,9155251,211253,Haider2023AFP,11044470}. Passive Near Field Communication (NFC) at $13.56~\mathrm{MHz}$ is particularly attractive because it is widely used for access control, identity verification, payment, and other user-verification services. Extending commercial NFC cards with an input capability could therefore allow users to enter a PIN directly on the same card used for identification. However, realizing such an interface requires reliable recognition of multiple digits, support for a complete set of control functions, and the ability to identify uncertain inputs that require repetition. Developing this capability on unmodified commercial NFC cards presents the following challenges. 

The first challenge is hardware compatibility. Existing passive RF interaction systems often require custom chipless radio-frequency identification (RFID) tags, embedded resonators, printed conductive structures, modified substrates, or specialized RF measurement hardware~\cite{Shahid2020ChiplessRT,10.1145/3631442}. Such tag-side modifications prevent direct use of unmodified commercial NFC cards, while specialized RF measurement hardware can complicate integration with standard NFC readers. The second challenge is interface completeness and hardware robustness. Prior PIN entry approaches often provide only reduced input interfaces without a complete numerical keypad and control functions or assume limited hardware variation~\cite{Sarpong2026CADRECD,Chen2024CrossDeviceRF,Yang2024MitigatingRI}. A practical system must support a complete PIN pad while remaining robust under simultaneous card and reader variations. The last challenge is uncertainty-aware recognition. Existing systems are typically designed around a specific RF interaction mechanism and recognition pipeline~\cite{8737419,10.1145/2702123.2702178,8486346}, using variations in impedance, resonance, backscatter, or magnetic coupling to recognize user input~\cite{10.1145/2858036.2858249,10.1145/3332165.3347910,Haider2023AFP}. Their interaction procedures are therefore tightly coupled to the measured RF response and learning model, while prior work commonly reports fixed-model recognition accuracy under specific card--reader configurations~\cite{Sarpong2026CADRECD}. However, a practical system requires not only point recognition but also reliable quantification of recognition uncertainty to identify ambiguous inputs that require additional observations~\cite{Ovadia2019TrustUncertainty}. Conventional classifiers instead return a single label even when the RF evidence is ambiguous, without explicitly determining whether recognition is sufficiently reliable~\cite{pmlr-v235-huang24aa,pmlr-v235-kiyani24a}. A practical NFC PIN entry system should therefore not only recognize button classes but also determine when an input is sufficiently reliable for acceptance and when re-input is required to obtain additional evidence for a reliable decision.    

\begin{figure}[!t]
    \centering
    \includegraphics[width=\columnwidth]{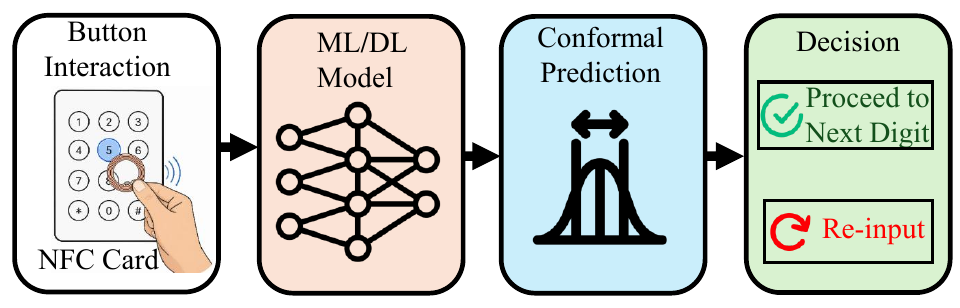}
    \caption{Illustration of the proposed adaptive virtual NFC PIN entry framework.}
    \label{fig:system_overview}
    \vspace{-10pt}
\end{figure}

In this paper, we use a commercial NFC ID card as the interaction surface for passive user input. The card is divided into 12 uniform \textit{virtual buttons}, forming a personal \textit{virtual PIN pad} with ten numerical digits (0--9) and two control buttons for confirmation and deletion. A user selects a button by placing a small passive resonant coil, such as an NFC key fob, over the corresponding location. The coil perturbs the magnetic coupling among the reader antenna, NFC card antenna, and button coil, producing a location-dependent RF response without modifying the card or the underlying ISO/IEC~15693 communication protocol.

To reliably recognize these inputs, we develop an adaptive and protocol-transparent virtual NFC input framework that addresses simultaneous card and reader shifts to recognize button input. Specifically, multi-target domain adaptation reduces card- and reader-dependent variations while preserving button-discriminative information. In addition, a passive analog self-interference cancellation (SIC) circuit suppresses transmitted-carrier leakage and improves acquisition of the weak load-modulated card response. To further improve recognition reliability, we introduce a model-agnostic uncertainty quantification protocol based on conformal prediction (CP) that combines evidence across interactions, determines whether each recognition is sufficiently reliable for acceptance, and requests re-input only when the evidence is insufficient. Once an input is accepted, the reader provides feedback indicating that the user may proceed to the next PIN button. Importantly, the CP-based adaptive decision process requires no modification to the underlying trained deep learning (DL)/machine learning (ML) model or ISO/IEC~15693 protocol. For experimental validation, we further develop a fully reconfigurable software-defined radio (SDR)-based NFC reader compliant with ISO/IEC~15693. The testbed integrates passive analog SIC, 35 commercial NFC cards, and three reader configurations to evaluate cross-card, cross-reader, and combined card--reader domain shifts. Fig.~\ref{fig:system_overview} illustrates the overall workflow.

Generally, the contributions of this paper can be summarized as follows: 1) we develop a practical 12-button \textit{virtual PIN pad} using unmodified passive commercial NFC cards; 2) we develop a multi-target domain adaptation framework for DL and ML models that learns button-discriminative representations robust to cross-card, cross-reader, and simultaneous card--reader variations; 3) we develop a model-agnostic adaptive PIN-button entry protocol that uses uncertainty quantification to accept reliable inputs immediately, request re-input only when necessary, and provide acceptance or repetition feedback; and 4) we construct a fully reconfigurable SDR-based ISO/IEC~15693 testbed with passive analog SIC, multiple commercial NFC cards, and multiple reader configurations. The implementation code and resources are shared in~\cite{VirtualPINPad2026}.

% In our previous work, we presented CADRE~\cite{Sarpong2026CADRECD}, a domain adaptation and Mahalanobis distance-based recognition framework that improved virtual PIN-button recognition on passive NFC cards under cross-card variations. Building on this foundation, this paper extends CADRE by introducing a model-agnostic uncertainty quantification framework based on conformal prediction to enable reliable adaptive PIN-button entry. The proposed framework determines when a recognized input can be accepted and when additional input is required. Extensive experiments further validate the effectiveness of the proposed approach across different card and reader domains. This study focuses on individual virtual-button recognition and adaptive re-input decisions. Complete PIN-sequence evaluation and end-to-end authentication are beyond the scope of this study.

The rest of the paper is organized as follows. Section~\ref{sec:related_work} reviews related work on passive RF interaction, domain adaptation, and reliable recognition. Section~\ref{sec:physical_model} presents the physical modeling of virtual button interaction on passive NFC cards. Section~\ref{Data_collection} describes the system implementation, data collection, domain adaptation, and uncertainty quantification framework. Section~\ref{Evaluation} presents the experimental results and performance evaluation. Finally, Section~\ref{Conclude} concludes the paper.

\section{Related Work} 
\label{sec:related_work}
In this section, we review three complementary areas: passive RF interaction, cross-domain RF-based recognition, and methods for improving recognition reliability. Existing studies typically address these areas separately and do not integrate them into a unified and adaptive framework for standard-compliant passive NFC interaction.

\textbf{Passive RF-based interaction.}
Existing studies use passive RFID tags and chipless RF structures to support battery-free touch, motion, and discrete input. Prior work has shown that fingertip interaction with RFID tags perturbs the backscattered response and produces distinctive phase patterns that can be used for recognition~\cite{8737419}. In~\cite{10.1145/2702123.2702178}, passive ultra-high-frequency (UHF) RFID measurements are used to recognize touch and object motion from changes in received signal strength, phase, and tag-read rate. In~\cite{10.1145/2858036.2858249,211253,10.1145/3332165.3347910}, conductive traces, printed chipless Wi-Fi structures, and multiple RFID chips are used to create battery-free paper interfaces, touch points, and modular touch pads. KeyStub~\cite{10.1145/3631442} enables discrete keypad input by interfacing commodity RFID integrated circuits with multiple microwave-band resonant stubs, each designed to produce a distinct impedance perturbation. In~\cite{Shahid2020ChiplessRT,Haider2023AFP}, chipless RFID tags use dedicated resonators to recognize touch locations from changes in resonant-frequency or radar-cross-section responses. These systems demonstrate the feasibility of battery-free RF interaction but generally require modified tags, multiple RFID chips, printed conductive structures, or specially fabricated resonant elements.

\textbf{Cross-domain RF-based recognition.}
In recent years, RF recognition models have been shown to be sensitive to variations in wireless channels, devices, and signal-acquisition hardware. In~\cite{9155259}, wireless-channel variations are shown to obscure hardware-dependent RF features and degrade recognition performance. In~\cite{10034841,10345732}, contrastive learning, subdomain adaptation, adversarial learning, and collaborative inference are used to improve recognition across different receivers. In~\cite{Chen2024CrossDeviceRF,Yang2024MitigatingRI,11044651}, domain adaptation methods align signal distributions from different receiving devices to improve cross-device RF fingerprint recognition. In~\cite{10923619}, multisource domain adaptation improves cross-domain Wi-Fi activity recognition under varying deployment conditions. These methods demonstrate the value of cross-domain learning but mainly address active-transmitter identification or Wi-Fi sensing rather than passive RF-based/NFC interaction under cross-card, cross-reader, and combined card--reader variations.

\textbf{Reliable RF-based interaction recognition.}
Conventional classifiers return a single label for every input, even when the available signal evidence is insufficient or multiple classes remain possible. Repeated observations can improve recognition reliability through decision aggregation. Majority voting combines predicted labels~\cite{Kuncheva2014}, while probability averaging combines class probabilities across observations~\cite{667881}. Confidence-based early classification further considers prediction reliability and observation cost when determining whether additional measurements are needed~\cite{9564134}. However, these methods do not inherently provide calibrated prediction sets. Conformal prediction constructs calibrated prediction sets that contain the correct label with a user-specified marginal coverage level under appropriate exchangeability assumptions~\cite{3495724.3496026}. In~\cite{pmlr-v235-huang24aa,pmlr-v235-kiyani24a}, label-ranking scores and learned feature representations are introduced to improve prediction set efficiency and conditional behavior. These studies primarily focus on constructing calibrated prediction sets. They do not consider how the resulting uncertainty information can be adapted to control a physical interaction process, determine when another user input interaction is required, or combine evidence across successive interactions.

Addressing these limitations requires both robustness to hardware variations and uncertainty-aware interaction decisions. In our previous work, we presented CADRE~\cite{Sarpong2026CADRECD}, a domain adaptation and Mahalanobis distance-based recognition framework that improved virtual PIN-button recognition on passive NFC cards under cross-card variations. Building on this foundation, this paper extends CADRE by introducing a model-agnostic uncertainty quantification framework based on conformal prediction to enable reliable adaptive PIN-button entry. The proposed framework determines when a recognized input can be accepted and when additional input is required. Extensive experiments further validate the effectiveness of the proposed approach across different card and reader domains. This study focuses on individual virtual-button recognition and adaptive re-input decisions. Complete PIN-sequence evaluation and end-to-end authentication are beyond the scope of this study.

\section{Physical Modeling of Virtual Button Interaction on NFC Cards}
\label{sec:physical_model}
The proposed NFC virtual PIN pad consists of an NFC reader, a passive NFC card, and one movable passive button coil. The card surface is divided into twelve predefined regions, denoted by
$\mathcal{B}=\{\mathrm{B0},\mathrm{B1},\ldots,\mathrm{B11}\}$,
where $\mathcal{B}$ is the set of virtual buttons and each region represents one virtual button. These regions do not contain separate physical switches or dedicated coils. Instead, the same passive button coil is placed over different regions to select the corresponding input. Since the reader and card remain fixed, each button-coil position establishes a different magnetic-coupling condition and produces a corresponding variation in the received response.

\begin{figure}[t]
    \centering
    \includegraphics[width=0.90\linewidth]{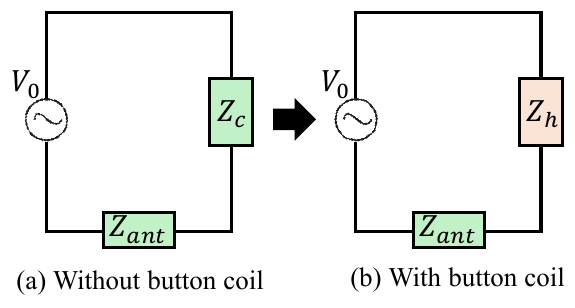}
    \caption{Equivalent circuit representation of the reader--card system with and without the passive button coil.}
    \label{fig:equivalent_circuit}
    \vspace{-10pt}
\end{figure}

\subsection{Button-Induced Reflected Impedance}
The reader, card, and button coils interact through near-field magnetic coupling. For two arbitrary coils indexed by $t$ and $q$, their mutual inductance is $M_{tq}$. For the remainder of this section, the subscripts $1$, $2$, and $b$ refer to the reader antenna, NFC card antenna, and passive button coil, respectively. Their impedances are denoted by $Z_{\mathrm{ant}}$, $Z_c$, and $Z_b$, respectively. Here, $Z_c$ includes the NFC card antenna, tuning network, and integrated-circuit load before the button coil is introduced, while $Z_b$ represents the impedance of the passive button coil. As illustrated in Fig.~\ref{fig:equivalent_circuit}, when the passive button coil is placed over a virtual button region, it magnetically couples with the NFC card antenna. Let $M_{2b}$ be the mutual inductance between the NFC card antenna and the passive button coil. The effective card impedance after the button coil is introduced is approximated as
$Z_h=Z_c+\frac{\omega^2M_{2b}^{2}}{Z_b}$,
where $Z_h$ denotes the button-perturbed effective card impedance, $\omega=2\pi f_c$ is the angular carrier frequency, and $f_c$ is the NFC carrier frequency. The first term represents the card impedance, while the second represents the impedance reflected from the button coil into the card circuit. Therefore, each button position produces a different card-button coupling coefficient $k_{2b}$ and a corresponding change in $Z_h$.

The effective card impedance and the direct reader-button coupling both contribute loading to the reader. The resulting reflected impedance is approximated as~\cite{10.1145/3714394.3754366}
\begin{equation}
Z_{\mathrm{ref}}
=
\frac{\omega^2M_{12}^{2}}{Z_h}
+
\frac{\omega^2M_{1b}^{2}}{Z_b},
\label{eq:reflected_impedance}
\end{equation}
where $Z_{\mathrm{ref}}$ denotes the total impedance reflected into the reader antenna, $M_{12}$ denotes the mutual inductance between the reader and NFC card antennas, and $M_{1b}$ denotes the mutual inductance between the reader antenna and the passive button coil. The first term represents the card loading, including the button-induced perturbation contained in $Z_h$, while the second represents the direct loading introduced by the button coil.

The total impedance observed at the reader is
$Z_{\mathrm{in}}=Z_{\mathrm{ant}}+Z_{\mathrm{ref}}$,
where $Z_{\mathrm{in}}$ denotes the input impedance observed at the reader. For an RF interface with reference impedance $Z_0$, the corresponding reflection coefficient is
\begin{equation}
\Gamma
=
\frac{Z_{\mathrm{in}}-Z_0}
{Z_{\mathrm{in}}+Z_0},
\label{eq:reflection_coefficient}
\end{equation}
where $\Gamma$ denotes the reader-side reflection coefficient and $Z_0$ denotes the reference impedance of the RF interface. Thus, moving the button coil changes $M_{2b}$, $k_{2b}$, $Z_h$, and $Z_{\mathrm{ref}}$, producing repeatable variations in the magnitude and phase of $\Gamma$.

\begin{figure}[t]
    \centering
    \includegraphics[width=0.90\linewidth]{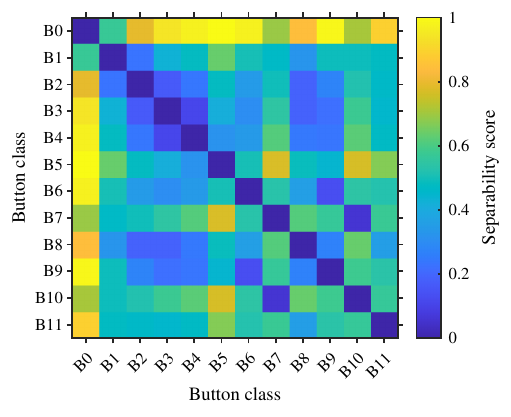}
    \caption{Impact of button coil position on virtual button class separability. }
    \label{fig:Button_Separability_Comsol}
    \vspace{-15pt}
\end{figure} 

Next, we evaluate the effect of button-coil position using the RF Module in COMSOL Multiphysics. The model consists of a three-turn rectangular NFC card antenna and a passive button coil positioned \(5.5~\mathrm{mm}\) above the card surface. The card is excited over a frequency range around \(13.56~\mathrm{MHz}\), and the same button coil is placed at each of the twelve virtual button regions, B0--B11, while its geometry, orientation, material properties, and vertical separation remain fixed. The regions form a \(3\times4\) grid with column coordinates \(x=\{-30,0,30\}~\mathrm{mm}\) and row coordinates \(y=\{20,6.7,-6.7,-20\}~\mathrm{mm}\), measured relative to the card center. For button \(b\), \(\Gamma_b(f)\) denotes the complex reflection coefficient with the button coil present, while \(\Gamma_0(f)\) denotes the card-only response. The button-induced complex perturbation is defined as \(\Delta\Gamma_b(f)=\Gamma_b(f)-\Gamma_0(f)\). Its frequency-dependent shape is characterized using the numerical derivative \(g_b(f)=\partial|\Delta\Gamma_b(f)|/\partial f\), evaluated from the discrete COMSOL frequency samples using finite differences. For each button pair \((b,q)\), the slope difference at frequency \(f_n\) is computed as \(D_{bq}(f_n)=|g_b(f_n)-g_q(f_n)|\). These values are normalized according to \(\widetilde{D}_{bq}(f_n)=D_{bq}(f_n)/D_{\max}\), where \(D_{\max}=\max_{b,q,n}D_{bq}(f_n)\). The pairwise RF-response separability score is then obtained as \(S_{bq}=(1/N_f)\sum_{n=1}^{N_f}\widetilde{D}_{bq}(f_n)\), where \(N_f\) is the number of sampled frequencies. Thus, \(S_{bb}=0\), \(S_{bq}=S_{qb}\), and \(0\leq S_{bq}\leq1\), with larger values indicating greater differences between the frequency-dependent response shapes of two button positions. As shown in Fig.~\ref{fig:Button_Separability_Comsol}, the nonzero off-diagonal scores demonstrate that changing only the lateral position of the same passive coil produces distinct reflection-coefficient responses. The unequal scores also show that some button pairs are more difficult to distinguish, motivating the subsequent learning and uncertainty-aware recognition framework.

\section{System Design and Implementation}
\label{Data_collection}
In this section, we present the experimental prototype and learning framework for reliable virtual PIN-pad button recognition under multiple domain shifts. We describe the prototype, hardware-effect mitigation, ISO/IEC~15693 communication protocol, and multi-card, multi-reader dataset collection and preprocessing. We then introduce the DL and Random Forest models to demonstrate feasibility across both model types, multi-domain adaptation for robustness to unseen cards, readers, and combined shifts, and conformal prediction for uncertainty-aware requests for additional inputs. All communication follows ISO/IEC~15693~\cite{iso15a}, where the NFC cards are vicinity integrated circuit cards (VICCs) containing 12 virtual button regions, and the NFC reader is the vicinity coupling device (VCD).

\begin{figure}[t]
    \centering
    \includegraphics[width=\columnwidth]{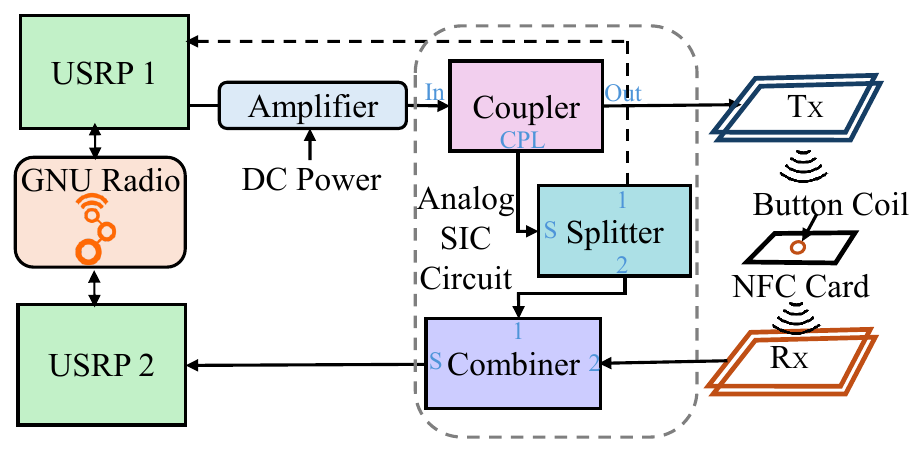}
    \caption{Block diagram of the SDR-based NFC testbed.}
    \label{fig:Block_Diagram_Testbed}
\end{figure}

\begin{figure}[t]
    \centering
    \includegraphics[width=\columnwidth]{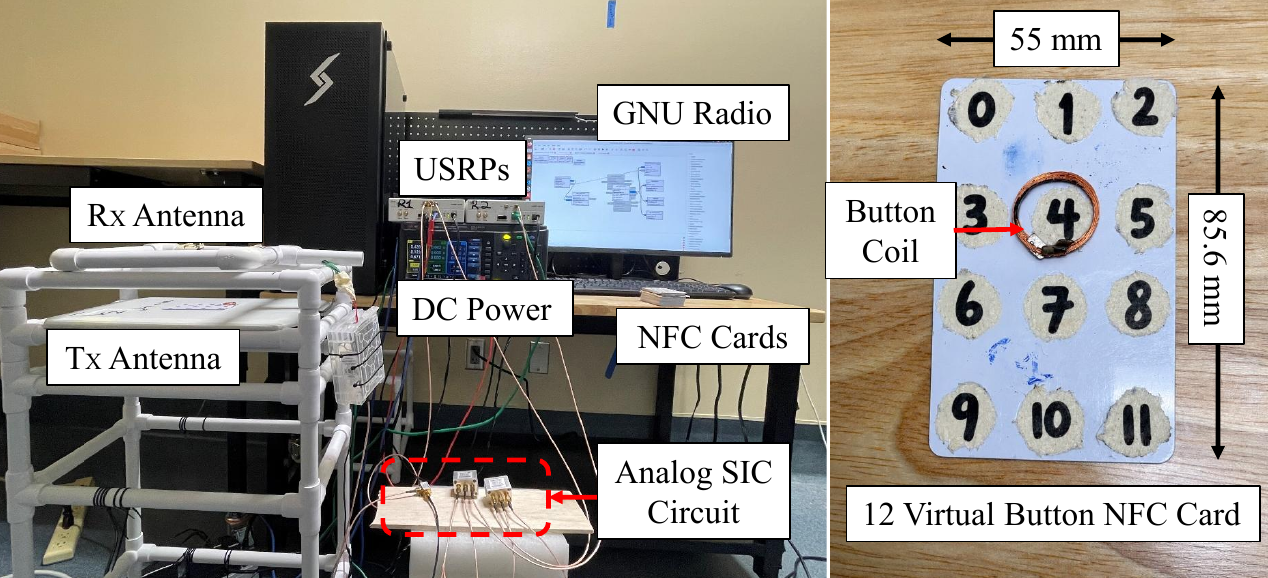}
    \caption{Prototype setup and virtual PIN pad NFC card.}
    \label{fig:experimental_testbed}
    \vspace{-15pt}
\end{figure}

\subsection{Prototype Implementation and Data Collection} 
\label{subsec:prototype_data_collection}
We build an ISO/IEC~15693-compliant testbed to collect the virtual PIN pad dataset. Fig.~\ref{fig:Block_Diagram_Testbed} shows the block diagram of the SDR-based NFC testbed, while Fig.~\ref{fig:experimental_testbed} presents the corresponding hardware implementation. The prototype consists of a software-defined NFC reader, separate transmit and receive RF paths, a passive analog self-interference cancellation circuit, custom receive antennas, and a GNU Radio-based signal-acquisition pipeline. This configuration enables the passive NFC card to be continuously powered while its weak load-modulated response is simultaneously captured. It also supports controlled measurements across different virtual button locations, NFC cards, and reader configurations.

\textbf{Transmit and receive antennas:}
The reader uses two Ettus Research USRP N210 devices to separate signal transmission and reception. USRP~1 generates the reader commands and is connected through the transmit RF chain to a FEIG ISC.ANT340/240-A loop antenna. The antenna operates at the standard NFC carrier frequency of $13.56~\mathrm{MHz}$ and generates the magnetic field required to power and communicate with the passive NFC card. USRP~1 transmits the reader commands using $100\%$ amplitude-shift keying and the ISO/IEC~15693 1-out-of-4 pulse-position coding scheme. USRP~2 captures the card response through the receive RF path and the analog self-interference cancellation circuit. Because the card response is substantially weaker than the transmitted carrier, we construct three customized three-turn rectangular receive antennas using AWG~14 copper wire with a cross-sectional area of approximately $2.08~\mathrm{mm}^{2}$. Each receive antenna's dimensions are approximately $200~\mathrm{mm}\times240~\mathrm{mm}$. The measured self-inductance of each receive antenna is approximately $6.5~\mu\mathrm{H}$ at $1~\mathrm{MHz}$. Each receive antenna is connected to a tuning network consisting of an approximately $1.5~\mu\mathrm{H}$ series inductor, a $33~\mathrm{pF}$ fixed capacitor, and a series-connected tunable capacitor. The tunable capacitor is adjusted separately for each antenna until its resonance is located near $14.01~\mathrm{MHz}$, close to the upper sideband of the ISO/IEC~15693 load-modulated response. Antenna resonance is measured using a RIGOL RSA3015N operating in vector network analysis mode after one-port open--short--load calibration using a RIGOL CK106E calibration kit. This frequency-selective tuning reduces the relative contribution of the strong $13.56~\mathrm{MHz}$ carrier and improves sensitivity to the weaker card-response sideband. The three independently constructed and tuned receive antennas are used to establish the different reader configurations considered in the cross-reader evaluation.

\textbf{Passive analog self-interference cancellation:}
The continuous-wave (CW) carrier signal transmitted by USRP~1 is significantly stronger than the load-modulated response generated by the passive card. A portion of this carrier couples directly from the transmit antenna and RF chain into the receive antenna. Without suppression, this leakage can dominate the received waveform, reduce the available receiver dynamic range, and potentially saturate USRP~2. To reduce the leakage before signal digitization and processing, we implement a passive analog self-interference cancellation circuit using one Mini-Circuits ZX30-9-4-S+ $8.9$-dB directional coupler and two Mini-Circuits ZMSC-2-2B two-way power splitter/combiners. As shown in Fig.~\ref{fig:Block_Diagram_Testbed}, the amplified signal from USRP~1 is applied to the \textit{In} port of the directional coupler. The \textit{Out} port feeds the FEIG transmit antenna, while the \textit{CPL} port provides an attenuated reference copy of the transmitted carrier. The reference and receive branches experience different attenuation and phase responses through the coupler, splitter, interconnecting cables, antennas, and RF components. When combined, the resulting amplitude and phase relationships partially suppress the directly coupled carrier before the signal reaches USRP~2. Because these relationships vary with the antenna configuration and surrounding environment, some residual carrier leakage remains at the combiner output. The receive antenna tuning further reduces the residual carrier relative to the upper load-modulation sideband. Together, the tuned receive antenna and passive cancellation circuit improve the usable receiver dynamic range and support reliable acquisition of button responses.

\textbf{NFC card virtual PIN pad configuration:} A button coil is placed at designated locations on a commercial NXP ICODE SLIX card operating at $13.56~\mathrm{MHz}$~\cite{nxpicode2018}. The $85.6~\mathrm{mm}\times55.0~\mathrm{mm}$ card surface is divided into 12 virtual buttons arranged as a $3\times4$ grid, with each region measuring approximately $28.5~\mathrm{mm}\times13.8~\mathrm{mm}$. The column centers are located at $x\in\{-28.5,0,+28.5\}~\mathrm{mm}$ and the row centers at $y\in\{+20.6,+6.9,-6.9,-20.6\}~\mathrm{mm}$. The buttons are labeled row-wise from B0 to B11. During data collection, the coil is centered within each region to ensure consistent reference positions, while controlled placement variations are introduced to represent practical use.

\textbf{ISO/IEC 15693 communication:}
Communication between the reader and the NFC cards is performed according to the ISO/IEC~15693 inventory procedure~\cite{iso15a}. Each request includes a start-of-frame (SOF), a Flags field, an Inventory command, a mask-length field, a cyclic redundancy check (CRC), and an end-of-frame (EOF). The mask length is set to zero, while the optional application family identifier (AFI) is not included.  The card response follows the one-subcarrier low-data-rate format specified in ~\cite{iso15a}, using a $423.75~\mathrm{kHz}$ subcarrier and a data rate of approximately $6.62~\mathrm{kbit/s}$. This mode uses longer symbols and a narrowband response, yielding stable signals for repeatable acquisition~\cite{Sarpong2025ModelAgnosticUQ}.

\textbf{Signal acquisition and preprocessing:}
The data-collection software is implemented in GNU Radio. Reader commands are generated at $12.5~\mathrm{MS/s}$, while the complex card responses are recorded at $2~\mathrm{MS/s}$. The expected response window is isolated using the known command duration. A correlation-based detector locates the beginning of the card response, after which a fixed-length segment is extracted for recognition. We retain the first eight response bits because they remain constant across cards and do not depend on the variable payload. Slow signal drift is removed, and each extracted response is normalized before model training. For each card--reader collection, we collect 200 valid responses from each of the 12 virtual buttons. Controlled button coil placement variations are introduced after every 50 responses. Each card--reader collection therefore contains $2{,}400$ samples.

We use three reader configurations, denoted by $\mathcal{R}_1$, $\mathcal{R}_2$, and $\mathcal{R}_3$. Specifically, $\mathcal{R}_1=\{\mathrm{Tx}_1,\mathrm{Rx}_1\}$, $\mathcal{R}_2=\{\mathrm{Tx}_2,\mathrm{Rx}_2\}$, and $\mathcal{R}_3=\{\mathrm{Tx}_3,\mathrm{Rx}_3\}$, where $\mathrm{Tx}_i$ and $\mathrm{Rx}_i$ denote the transmit and receive antennas used in the $i$th reader configuration, respectively. Each configuration therefore uses a different FEIG transmit antenna together with a separately constructed and tuned receive antenna, allowing the evaluation to capture variations in reader-side antenna and tuning configurations. The domain adaptation data are organized into four domains:
\begin{enumerate}
    \item \textbf{Source Domain:} Cards~1--5 are collected using reader configuration $\mathcal{R}_1$.
    \item \textbf{Cross-Card (CC):} Cards~6--10 are collected using reader configuration $\mathcal{R}_1$. The reader configuration remains unchanged, while the card identities differ from those in the source domain.
    \item \textbf{Cross-Reader (CR):} The same source Cards~1--5 are collected using reader configuration $\mathcal{R}_2$. The card identities remain unchanged, while both the transmit and receive antenna configurations differ from those in the source domain.
    \item \textbf{Cross-Card-and-Reader (CCR):} Cards~11--15 are collected using reader configuration $\mathcal{R}_2$. Therefore, both the card identities and the reader configuration differ from those in the source domain.
\end{enumerate}
The domain adaptation dataset contains 20 card--reader collections and $48{,}000$ response samples. We additionally collect a separate evaluation dataset, subsequently divided into calibration, validation, and held-out test subsets, to assess the proposed adaptive recognition framework under the three considered domain shifts. The CC evaluation set contains Cards~16--25 collected using reader configuration $\mathcal{R}_1$, resulting in $24{,}000$ responses. The CR evaluation set contains the five source-domain cards, i.e., Cards~1--5, collected using the previously unseen reader configuration $\mathcal{R}_3$, resulting in $12{,}000$ responses. The CCR evaluation set contains Cards~26--35 collected using reader configuration $\mathcal{R}_3$, resulting in $24{,}000$ responses. Therefore, the evaluation dataset contains 25 card--reader collections and $60{,}000$ response samples. The complete dataset contains 45 card--reader collections and $108{,}000$ virtual-button response samples.

\subsection{Domain Adaptation Training}
We adapt and employ the Correlation Alignment (CORAL) framework~\cite{Sun2016DeepCC,8434290}~for the multi-target domain adaptation setting considered in this work. Specifically, for the DL model, we use a multi-target Deep CORAL objective that jointly optimizes source-button classification and feature alignment with the CC, CR, and CCR target domains to handle cross-card and cross-reader virtual-button recognition. For the ML model, CORAL is applied as a standalone feature-alignment step before classifier training. The source domain contains labeled virtual-button responses collected using the reference card--reader configuration. The source labels supervise button recognition, whereas the target responses contribute only to feature-distribution alignment. The labels of the target responses used for domain alignment are not used during training. Final evaluation is performed using separate held-out target responses and their corresponding labels.

\begin{figure}
    \centering
    \includegraphics[width=\linewidth]{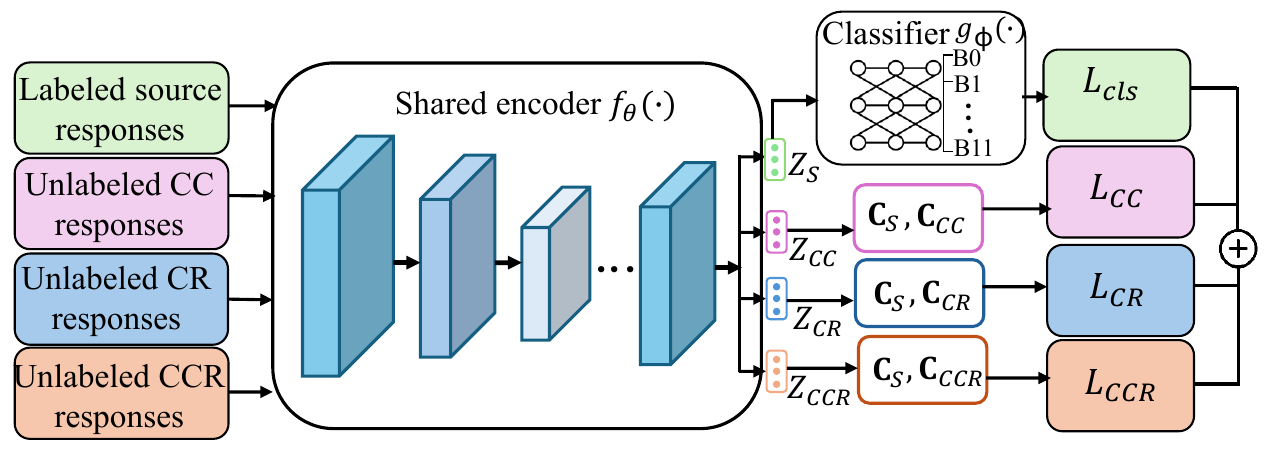}
    \caption{Multi-target domain adaptation architecture for the DL model.} 
    \label{fig:Multi-shift DG Architect}
    \vspace{-15pt}
\end{figure}

\subsubsection{Deep Learning Model}
As illustrated in Fig.~\ref{fig:Multi-shift DG Architect}, the DL model consists of a shared temporal encoder and a virtual-button classifier. The encoder processes responses from the source and all target domains using the same network parameters and maps them into a common feature space. The classifier is trained using only the labeled source responses, while the Deep CORAL objective aligns the second-order feature statistics of the source domain with those of the unlabeled CC, CR, and CCR domains. 

The shared encoder first applies two one-dimensional convolutional blocks with a kernel size of seven. The first convolutional layer contains 32 filters, while the second contains 64 filters. Each convolutional layer is followed by batch normalization, ReLU activation, and max pooling. The resulting temporal feature sequence is then processed by a 64-unit Long Short-Term Memory (LSTM) layer with sequence outputs retained. Global average pooling removes the temporal dimension, after which a fully connected layer with 64 units and ReLU activation produces the learned feature representation. Dropout is applied for regularization, resulting in a 64-dimensional encoder embedding. The classifier then applies an additional fully connected layer with 64 units, followed by ReLU activation, batch normalization, and dropout. Finally, a fully connected output layer produces twelve logits corresponding to the twelve virtual buttons, and the softmax function converts these logits into class probabilities when required. Let $f_{\theta}(\cdot)$ denote the shared encoder, where $\theta$ represents its trainable parameters. For an input response $x$, the encoder produces $z=f_{\theta}(x)$, where $z\in\mathbb{R}^{d}$ is the learned feature vector and $d=64$ is the feature dimension. Let $g_{\phi}(\cdot)$ denote the button classifier, where $\phi$ represents its trainable parameters. The probability that response $x$ corresponds to virtual button $\mathrm{B}_b$ is
\begin{equation}
p(\mathrm{B}_b\mid x)
=
\frac{
\exp\!\left(g_{\phi,b}(z)\right)
}{
\displaystyle
\sum_{\mathcal{u}=0}^{|\mathcal{B}|-1}
\exp\!\left(g_{\phi,\mathcal{u}}(z)\right)
},
\label{eq:button_probability}
\end{equation}
where $\mathcal{B}=\{\mathrm{B0},\mathrm{B1},\ldots,\mathrm{B11}\}$ is the set of virtual buttons, $b\in\{0,\ldots,11\}$ is the button index, $g_{\phi,b}(z)$ is the classifier logit associated with button $\mathrm{B}_b$, and $\mathcal{u}$ is the summation index in the softmax denominator.

At each training iteration, one labeled source mini-batch and one unlabeled mini-batch from each target domain are used. The labeled source mini-batch is represented as $\mathcal{M}_S=\{(x_n^S,y_n^S)\}_{n=1}^{N_S}$, where $x_n^S$ is the $n$th source response, $y_n^S$ is its corresponding button index, and $N_S$ is the source mini-batch size. For each target domain $D\in\mathcal{D}_T$, the corresponding unlabeled mini-batch is represented as $\mathcal{M}_D=\{x_n^D\}_{n=1}^{N_D}$, where $x_n^D$ is the $n$th response from target domain $D$, and $N_D$ is the corresponding target mini-batch size. Equal mini-batch sizes are used across the source and target domains to obtain covariance estimates with comparable statistical reliability. The labeled source mini-batch is used to optimize button recognition. Let $p_{n,b}^{S}$ denote the probability assigned to button $\mathrm{B}_b$ for source response $x_n^S$. The source classification loss is
\begin{equation}
L_{\mathrm{cls}}
=
-\frac{1}{N_S}
\sum_{n=1}^{N_S}
\log p_{n,y_n^S}^{S},
\label{eq:source_classification_loss}
\end{equation}
where $p_{n,y_n^S}^{S}$ denotes the predicted probability assigned to the true button class $y_n^S$ for source response $x_n^S$.

Deep CORAL incorporates second-order statistical alignment into the deep network by minimizing the batch-wise discrepancy between the covariance matrices of the source and target feature distributions~\cite{8434290}. For each target domain $D\in\mathcal{D}_T$, the source-to-target Deep CORAL loss is defined as
\begin{equation}
L_D
=
\frac{1}{4d^2}
\left\|
\mathbf{C}_S-\mathbf{C}_D
\right\|_F^2,
\label{eq:target_coral_loss}
\end{equation}
where the covariance matrix of the source features is $
\mathbf{C}_S
=
\frac{1}{N_S-1}
\left(
Z_S-\mathbf{1}_{N_S}\mu_S^T
\right)^T
\left(
Z_S-\mathbf{1}_{N_S}\mu_S^T
\right)
$, the covariance matrix of the target features is $
\mathbf{C}_D
=
\frac{1}{N_D-1}
\left(
Z_D-\mathbf{1}_{N_D}\mu_D^T
\right)^T
\left(
Z_D-\mathbf{1}_{N_D}\mu_D^T
\right)$, and $d$ is the feature dimension. The encoder outputs are arranged into the source feature matrix $Z_S\in\mathbb{R}^{N_S\times d}$ and the target-domain feature matrix $Z_D\in\mathbb{R}^{N_D\times d}$, where each row contains one feature vector $z_n^S=f_{\theta}(x_n^S)$ or $z_n^D=f_{\theta}(x_n^D)$. Their feature means are $\mu_S=\frac{1}{N_S}\sum_{n=1}^{N_S}z_n^S$ and $\mu_D=\frac{1}{N_D}\sum_{n=1}^{N_D}z_n^D$. The notation $(\cdot)^T$ denotes matrix transpose, $\mathbf{1}_{N_S}$ and $\mathbf{1}_{N_D}$ are all-one column vectors, and $\|\cdot\|_F$ denotes the Frobenius norm. Minimizing $L_D$ reduces the mismatch between the second-order feature statistics of the source and target domains, while the source classification objective encourages the learned representation to remain discriminative for the virtual buttons. Although the feature means are used to center the samples when computing the covariance matrices, the CORAL loss does not explicitly penalize the difference between the source and target feature means. In the adopted architecture, batch normalization and joint optimization with the source classification objective help stabilize the shared feature representation, but they do not constitute an explicit mean-alignment constraint. Because balanced mini-batches are used across CC, CR, and CCR, the three target-specific alignment losses are uniformly averaged so that no target domain dominates the adaptation objective. This design favors a single general model across the considered deployment conditions rather than assigning additional weight to a particular target domain.

The source classification loss and the averaged multi-target alignment loss are combined as
\begin{equation}
L_{\mathrm{total}}
=
L_{\mathrm{cls}}
+
\frac{\lambda_{\mathrm{DA}}}{N_T}
\sum_{D\in\mathcal{D}_T}
L_D,
\label{eq:multi_target_deep_coral}
\end{equation}
where $\lambda_{\mathrm{DA}}$ controls the contribution of domain alignment and $N_T$ is the number of target domains. The classification term encourages the learned representation to remain discriminative for the twelve virtual buttons, while the alignment term reduces the mismatch in card- and reader-dependent second-order feature statistics between the source and target domains. The encoder and classifier parameters are jointly optimized according to
\begin{equation}
(\theta^*,\phi^*)
=
\arg\min_{\theta,\phi}
L_{\mathrm{total}},
\label{eq:joint_deep_coral_optimization}
\end{equation}
where $\theta^*$ and $\phi^*$ denote the learned encoder and classifier parameters, respectively. During each training iteration, the labeled source responses and unlabeled responses from all target domains are passed through the shared encoder. The source features are forwarded to the classifier to compute $L_{\mathrm{cls}}$, while the source and target features are used to compute the three CORAL losses. These losses are averaged and combined with $L_{\mathrm{cls}}$ before performing a single backward pass. The gradient of $L_{\mathrm{cls}}$ propagates through both the classifier and the shared encoder, whereas the gradients of the CORAL losses propagate only through the encoder because the alignment losses are computed directly from the encoder features. All trainable parameters are updated using the Adam optimizer. Thus, source-button classification and multi-target feature alignment are performed within the same training loop. Although Deep CORAL aligns global second-order feature statistics rather than class-conditional distributions, the source classification objective encourages the shared encoder to preserve button-discriminative structure while suppressing domain-dependent variations. Since this global alignment does not explicitly guarantee class-conditional correspondence across domains, its effectiveness is evaluated empirically using held-out target-domain accuracy, confusion matrices, and feature visualizations.

At inference time, only the trained encoder and button classifier are required. For a target response $x^t$, the predicted virtual-button index is
\begin{equation}
\hat{b}
=
\arg\max_{b\in\{0,\ldots,11\}}
p(\mathrm{B}_b\mid x^t).
\label{eq:dl_button_prediction}  
\end{equation}
The labels of the target responses used for domain alignment are not used during adaptation. The domain-adaptation target labels are reserved exclusively for evaluating recognition performance.

\begin{figure*}[t]
    \centering
    % DL pair: left column
    \begin{minipage}[t]{0.49\textwidth}
        \centering
        \subfloat[\label{fig:tsne_after_dg}]{%
            \begin{minipage}[t]{0.50\linewidth}
                \centering
                \makebox[\linewidth][c]{%
                    \includegraphics[
                        width=1.08\linewidth,
                        trim=4pt 6pt 4pt 4pt,
                        clip
                    ]{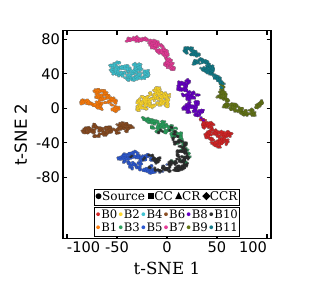}%
                }
            \end{minipage}%
        }%
        \subfloat[\label{fig:cm_after_dl}]{%
            \begin{minipage}[t]{0.50\linewidth}
                \centering
                \makebox[\linewidth][c]{%
                    \hspace*{-8pt}%
                    \includegraphics[
                        width=1.08\linewidth,
                        trim=4pt 6pt 4pt 4pt,
                        clip
                    ]{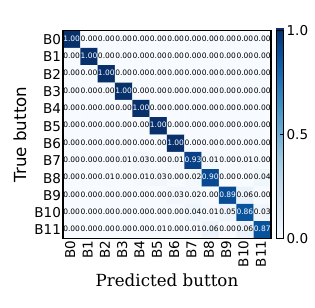}%
                }
            \end{minipage}%
        }
        \caption{DL model: (a) visualization of the learned feature
        distributions using t-SNE; (b) confusion matrix.}
        \label{fig:dl_features_confusion}
    \end{minipage}\hfill
    % RF pair: right column
    \begin{minipage}[t]{0.49\textwidth}
        \centering
        \subfloat[\label{fig:tsne_after_dg_rf}]{%
            \begin{minipage}[t]{0.50\linewidth}   
                \centering 
                \makebox[\linewidth][c]{%
                    \includegraphics[
                        width=1.08\linewidth,
                        trim=4pt 6pt 4pt 4pt,
                        clip
                    ]{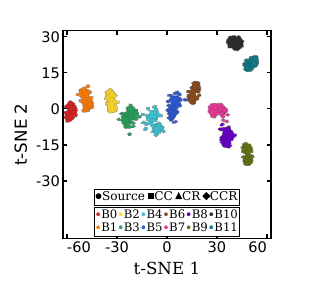}%
                }
            \end{minipage}%
        }%
        \subfloat[\label{fig:cm_after_rf}]{%
            \begin{minipage}[t]{0.50\linewidth}
                \centering
                \makebox[\linewidth][c]{%
                    \hspace*{-8pt}%
                    \includegraphics[
                        width=1.08\linewidth,
                        trim=4pt 6pt 4pt 4pt,
                        clip
                    ]{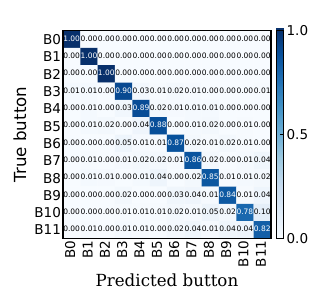}%
                }
            \end{minipage}%
        }
        \caption{Random Forest model: (a) visualization of the learned
        feature distributions using t-SNE; (b) confusion matrix.}
        \label{fig:rf_features_confusion}
    \end{minipage}
     \vspace{-5pt}
\end{figure*}
To better understand how the multi-target domain adaptation model generalizes across the considered card--reader conditions, we repeat the training and evaluation procedure across five card-based splits, with the results shown in Figs.~\ref{fig:tsne_after_dg} and~\ref{fig:cm_after_dl}. After adaptation, responses corresponding to the same virtual button form compact clusters across the source, CC, CR, and CCR domains, as shown in Fig.~\ref{fig:tsne_after_dg}. The confusion matrices similarly show strong diagonal concentration and few off-diagonal errors after adaptation. Across the five splits, domain adaptation improves the average test accuracy on the combined CC, CR, and CCR data from $93.75\%\pm0.42\%$ to $95.42\%\pm0.29\%$. After adaptation, the trained model is saved and subsequently used for virtual-button recognition and CP-based adaptive decision-making.

\subsubsection{Random Forest Model}
We also employ a classical ML model, namely a Random Forest classifier, for virtual-button recognition and show the model-agnostic property of the proposed solution. Since Random Forest is not differentiable, CORAL is applied as a standalone feature-alignment step before classifier training using the implementation provided by the ADAPT Python library~\cite{de2021adapt}. The button responses are first projected into a lower-dimensional feature space using principal component analysis (PCA). PCA is fitted using only the labeled source training responses and is subsequently applied to the source and all unlabeled target-domain responses. A single Random Forest classifier is trained using a common target representation estimated from the CC, CR, and CCR domains. Let $\mu_D^P$ and $\mathbf{C}_D^P$ denote the mean vector and covariance matrix of the PCA feature distribution from target domain $D\in\mathcal{D}_T$. The common target mean is computed as $\overline{\mu}_T^P=\frac{1}{N_T}\sum_{D\in\mathcal{D}_T}\mu_D^P,$ while the common target covariance matrix is computed as $\overline{\mathbf{C}}_T^P=\frac{1}{N_T}\sum_{D\in\mathcal{D}_T}\mathbf{C}_D^P$. Since balanced target datasets are used, the CC, CR, and CCR domains contribute equally to the common target statistics. The multi-target CORAL linear transformation is computed as~\cite{Sun2016CorrelationAF} $A_{\mathrm{MT}}=\left(\mathbf{C}_S^P+\epsilon_{\mathrm{RF}}I\right)^{-1/2}\left(\overline{\mathbf{C}}_T^P+\epsilon_{\mathrm{RF}}I\right)^{1/2},$ where $\mathbf{C}_S^P$ is the covariance matrix of the source PCA feature distribution, $I$ is the identity matrix, and $\epsilon_{\mathrm{RF}}$ is a small positive regularization value used to improve numerical stability. The first term whitens the source PCA features by removing their feature correlations, while the second term re-colors the whitened features using the common covariance estimated from the unlabeled target domains. The aligned source features are obtained as
\begin{equation}
\widetilde{H}_S
=
\left(H_S-\mathbf{1}_{N_S}(\mu_S^P)^T\right)
A_{\mathrm{MT}}
+
\mathbf{1}_{N_S}
(\overline{\mu}_T^P)^T, 
\end{equation}
where $H_S$ is the source PCA feature matrix and $\mu_S^P$ is its mean vector. This transformation maps the labeled source features into the common target feature space and reduces the mismatch between their second-order statistics and those of the unlabeled CC, CR, and CCR feature distributions.

The transformed source features retain their original button labels and are used to train the Random Forest classifier. The model uses 300 decision trees and produces a probability for each of the twelve virtual buttons by averaging the class probabilities generated by the individual trees. At inference time, each target response is projected using the same source-fitted PCA transformation and evaluated directly by the trained Random Forest classifier. Since the source features are aligned to the estimated common target representation during training, no additional CORAL transformation is applied to the target responses during inference. Fig.~\ref{fig:tsne_after_dg_rf} illustrates the resulting button-specific feature distributions across the source, CC, CR, and CCR domains.

To further evaluate the effect of domain adaptation on the Random Forest model, we obtained test results from five independent card-based evaluation runs. In each run, the complete Random Forest training and evaluation procedure is repeated using the same source-training and multi-target adaptation data. Performance is evaluated on the combined CC, CR, and CCR test responses, and the reported mean and standard deviation are computed across the five runs. Across these evaluations, domain adaptation improves the average Random Forest test accuracy from $85.00\%\pm1.12\%$ to $89.11\%\pm2.89\%$. The corresponding confusion matrix in Fig.~\ref{fig:cm_after_rf} illustrates the per-button recognition performance after adaptation. After adaptation, the trained Random Forest model is saved and subsequently used for virtual-button recognition and CP-based adaptive decision-making.

\subsection{Adaptive PIN-Button Recognition} 
\label{subsec:cp_adaptive_recognition}
Next, we employ a model-agnostic CP framework to quantify the uncertainty of virtual-button PIN-button recognition and improve decision reliability~\cite{10.1561/2200000101}. Conventional recognition models, such as classical ML models, select the button having the highest output score among the 12 virtual-button classes. However, the magnitude of this score is model dependent and, by itself, does not provide a statistically calibrated indication that the selected label is reliable~\cite{Ovadia2019TrustUncertainty}. To address this limitation, CP is applied after the underlying recognition ML/DL model has been trained. Let $M_m$ denote the selected pretrained ML/DL recognition model. CP uses a labeled calibration set $\mathcal{X}_{\mathrm{cal}}$, whose responses are used only to determine the conformal quantiles and are not used to train the ML/DL models. We implement CP calibration using the open-source MAPIE library with the pretrained ML/DL models~\cite{CORDIER}. For each CP evaluation domain, labeled responses from the separate final CP evaluation dataset are used for calibration. MAPIE then constructs prediction sets at risk level $\alpha$. Under exchangeability, the sets provide marginal coverage of $1-\alpha$, with $\alpha$ controlling the tradeoff between coverage and set size. A smaller $\alpha$ requests higher target coverage and generally produces larger prediction sets. In contrast, a larger $\alpha$ typically produces smaller prediction sets but permits a higher miscoverage probability. Accordingly, the prediction set size provides an operational measure of the recognition model's uncertainty or confidence. A single button prediction set provides concentrated support for one virtual button, whereas a larger prediction set indicates that the received response remains compatible with multiple button classes. Based on this uncertainty information, rather than immediately converting every prediction set into a single decision, we propose an adaptive PIN-button entry protocol that determines whether the available button response provides sufficient evidence for acceptance. For a received button response $x$, the CP prediction set generated using model $M_m$ is expressed as
\begin{equation}
\mathcal{U}_{m}(x)
=
\mathcal{C}_{m,\alpha}
\left(
M_{m}(x)
\mid
\mathcal{X}_{\mathrm{cal}}
\right),
\label{eq:cp_prediction_set}
\end{equation}   
where $\mathcal{C}_{m,\alpha}(\cdot)$ denotes the CP calibration procedure associated with $M_m$ at risk level $\alpha$, and
$\mathcal{U}_{m}(x)\subseteq\{0,\ldots,11\}$ contains the virtual-button labels considered plausible for $x$, while $\left|\mathcal{U}_{m}(x)\right|$ denotes the prediction set size for $x$. Because a single response may be affected by noise or button-coil placement variation, CP is combined with an adaptive re-input strategy that accumulates evidence across repeated captures of the same intended virtual button. Let $x_{\ell}$ denote the response collected at capture step $\ell$, and let
$\mathcal{U}_{m,\ell}=\mathcal{U}_{m}(x_{\ell})$ denote its corresponding prediction set. The first capture is processed as the initial input, while additional captures are processed only when a re-input is requested. Each capture is evaluated using the same CP calibrated pretrained recognition model.

\begin{table*}[t]
\centering
\caption{Comparison of standard and adaptive recognition performance on the final held-out test data.}
\label{tab:cp_adaptive_summary}
\renewcommand{\arraystretch}{1.05}
\setlength{\tabcolsep}{7pt}
\begin{tabular}{lc|ccc|cccc}
\toprule
\multicolumn{1}{c}{\textbf{Model}}
&
\multicolumn{1}{c|}{\textbf{Standard Recognition}}
&
\multicolumn{3}{c|}{\textbf{Adaptive Recognition Accuracy (\%)}}
&
\multicolumn{4}{c}{\textbf{Average CP-Adaptive Metrics}} \\
\cmidrule(lr){2-2}
\cmidrule(lr){3-5}
\cmidrule(lr){6-9}
&
\textbf{Accuracy (\%)}
&
\textbf{CC}
&
\textbf{CR}
&
\textbf{CCR}
&
$\boldsymbol{A_{\mathrm{C}}(\%)}$
&
$\boldsymbol{A_{\mathrm{R}}(\%)}$
&
$\boldsymbol{\overline{U}}$
&
$\boldsymbol{\overline{R}}$ \\
\midrule
DL
&
93.15
&
$97.40\pm0.19$
&
$97.42\pm1.11$
&
$95.74\pm1.18$
&
97.71
&
96.85
&
1.34
&
0.55
\\
Random Forest
&
86.73
&
$90.00\pm0.87$
&
$91.65\pm0.92$
&
$89.99\pm1.11$
&
94.06
&
90.55
&
2.35
&
1.03
\\
\bottomrule
\end{tabular}
\end{table*}

\begin{figure*}[!t]
    \centering

    \begin{subfigure}[t]{0.235\textwidth}
        \centering
        \includegraphics[width=\linewidth]{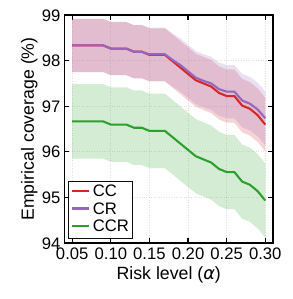}
        \caption{}
        \label{fig:empirical_coverage_dl}
    \end{subfigure}
    \hfill
    \begin{subfigure}[t]{0.235\textwidth}
        \centering
        \includegraphics[width=\linewidth]{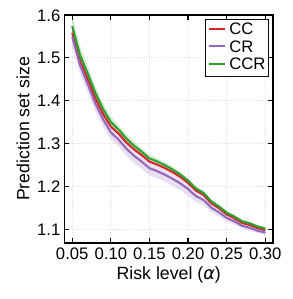}
        \caption{}
        \label{fig:prediction_set_size_dl}
    \end{subfigure}
    \hfill
    \begin{subfigure}[t]{0.235\textwidth}
        \centering
        \includegraphics[width=\linewidth]{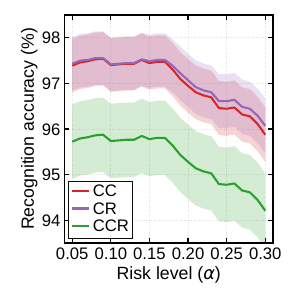}
        \caption{}
        \label{fig:adaptive_accuracy_dl}
    \end{subfigure}
    \hfill
    \begin{subfigure}[t]{0.235\textwidth}
        \centering
        \includegraphics[width=\linewidth]{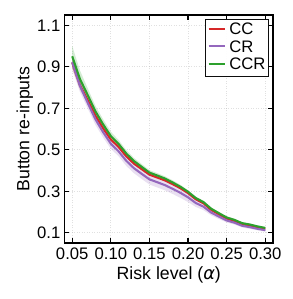}
        \caption{}
        \label{fig:average_reinputs_dl}
    \end{subfigure}

    \caption{Evaluation under varying risk levels using the DL model:
    (a) average empirical coverage,
    (b) average prediction set size,
    (c) recognition accuracy,
    and (d) average button re-inputs.}
    \label{fig:dl_adaptive_performance}
\end{figure*}

\begin{figure*}[!t]
    \centering

    \begin{subfigure}[t]{0.235\textwidth}
        \centering
        \includegraphics[width=\linewidth]{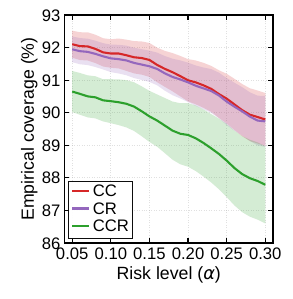}
        \caption{}
        \label{fig:empirical_coverage_rf}
    \end{subfigure}
    \hfill
    \begin{subfigure}[t]{0.235\textwidth}
        \centering
        \includegraphics[width=\linewidth]{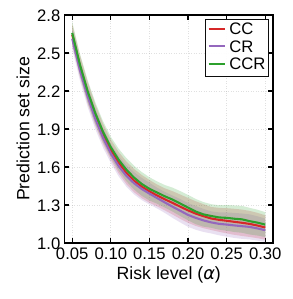}
        \caption{}
        \label{fig:prediction_set_size_rf}
    \end{subfigure}
    \hfill
    \begin{subfigure}[t]{0.235\textwidth}
        \centering
        \includegraphics[width=\linewidth]{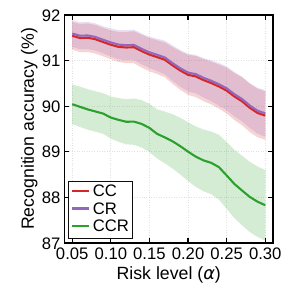}
        \caption{}
        \label{fig:adaptive_accuracy_rf}
    \end{subfigure}
    \hfill
    \begin{subfigure}[t]{0.235\textwidth}
        \centering
        \includegraphics[width=\linewidth]{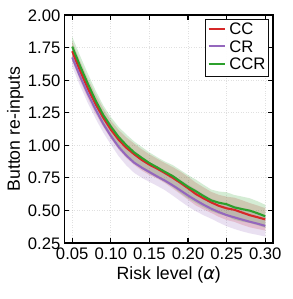}
        \caption{}
        \label{fig:average_reinputs_rf}
    \end{subfigure}

    \caption{Evaluation under varying risk levels using the Random Forest model:
    (a) average empirical coverage,
    (b) average prediction set size,
    (c) recognition accuracy,
    and (d) average button re-inputs.}
    \label{fig:rf_adaptive_performance}
     \vspace{-9pt}
\end{figure*}  

For each candidate button $b$, the system maintains a cumulative evidence score $c_b$. The complete score vector is
$\mathbf{c}=[c_0,c_1,\ldots,c_{11}]$, and all scores are initialized to zero before processing a new PIN-button input. After the first capture produces $\mathcal{U}_{m,1}$, the score assigned to Button~$b$ is
\begin{equation}
c_b=
\begin{cases}
\dfrac{1}{\left|\mathcal{U}_{m,1}\right|},
& \text{if } b\in\mathcal{U}_{m,1},\\[7pt]
0,
& \text{otherwise},
\end{cases}
\qquad
b\in\{0,\ldots,11\}.
\label{eq:initial_button_score}
\end{equation}
This rule assigns a total evidence mass of one to each nonempty prediction set and distributes that evidence uniformly among its included labels. Consequently, a singleton prediction set assigns one unit of evidence to its only candidate, whereas a prediction set containing several labels assigns a smaller contribution to each candidate. The update therefore accounts for both label inclusion and prediction set size. After each capture, the cumulative scores are compared with the empirically determined acceptance threshold $\tau$. If the accumulated score is insufficient, the reader immediately requests another input of the same intended virtual button. For each subsequent capture, the cumulative score of Button~$b$ is updated as
\begin{equation}
c_b=
\begin{cases}
c_b+\dfrac{1}{\left|\mathcal{U}_{m,\ell}\right|},
& \text{if } b\in\mathcal{U}_{m,\ell},\\[7pt]
c_b,
& \text{otherwise},
\end{cases}
\qquad
b\in\{0,\ldots,11\}.
\label{eq:cumulative_button_score}
\end{equation}
If an empty prediction set is produced, no cumulative score is updated for that capture, and the system requests another input when the interaction budget permits.

A button that appears consistently across successive prediction sets accumulates more evidence than labels that appear only occasionally. This reduces the influence of an isolated uncertain response and favors labels that are repeatedly supported across distinct captures. The cumulative evidence score is used as an operational decision statistic. After each capture, and until the maximum number of captures $L_{\max}$ is reached, the adaptive decision is determined by comparing the largest cumulative score with the acceptance threshold $\tau$. Let $c_{\max}=\max_{b\in\{0,\ldots,11\}} c_b.$ The recognized button $\hat{b}$ is 
\begin{equation}
\hat{b}
=
\begin{cases}
\displaystyle
\arg\max_{b\in\{0,\ldots,11\}} c_b,
& \text{if } c_{\max}\geq\tau,\\[6pt]
\text{request re-input},
& \text{if } c_{\max}<\tau
\text{ and } \ell<L_{\max},\\[6pt]
\displaystyle
\arg\max_{b\in\{0,\ldots,11\}} c_b,
& \text{if } c_{\max}<\tau
\text{ and } \ell=L_{\max}.
\end{cases}
\label{eq:adaptive_button_decision}
\end{equation}
When the first condition in~\eqref{eq:adaptive_button_decision} is satisfied, the button is accepted as the current PIN-button entry, the reader notifies the user, and the cumulative scores are reset for the next button. Otherwise, if $L_{\max}$ has not been reached, the reader requests repetition of the same intended button while retaining the cumulative scores. Acceptance occurs immediately after any capture that satisfies the threshold; if not completed by $L_{\max}$, the button with the highest cumulative evidence score is selected.

\section{Experimental Evaluation}
\label{Evaluation}
We use four main metrics to evaluate the performance of the proposed adaptive recognition system: adaptive recognition accuracy, average empirical coverage, average prediction set size, and average number of re-inputs. The adaptive recognition accuracy $A_{\mathrm{R}}$ is defined as the proportion of test instances that are correctly recognized after completing the adaptive decision procedure:
\begin{equation}
A_{\mathrm{R}}
=
\frac{1}{N_{\mathrm{test}}}
\sum_{i=1}^{N_{\mathrm{test}}}
\mathbb{I}
\left(
\hat{b}_i=b_i
\right),
\label{eq:adaptive_recognition_accuracy_eval}
\end{equation}
where $N_{\mathrm{test}}$ is the total number of test instances, $\hat{b}_i$ is the final virtual-button label produced for the $i$th test instance, $b_i$ is its corresponding ground-truth label, and $\mathbb{I}(\cdot)$ denotes the indicator function, which returns 1 when the final recognized label $\hat{b}_i$ matches the ground-truth label $b_i$ for the $i$th test instance and returns 0 otherwise. The average empirical coverage $A_{\mathrm{C}}$ is defined as the proportion of processed test responses for which the corresponding CP prediction set contains the ground-truth virtual-button label:
$A_{\mathrm{C}}=\frac{1}{N_{\mathrm{cap}}}\sum_{x_i\in\mathcal{X}_{\mathrm{cap}}}\mathbb{I}\left(b_i\in\mathcal{U}_{m}(x_i)\right)$,
where $\mathcal{X}_{\mathrm{cap}}$ contains all initial and re-input responses processed during testing, $N_{\mathrm{cap}}=|\mathcal{X}_{\mathrm{cap}}|=\sum_{i=1}^{N_{\mathrm{test}}}L_i$ is the total number of processed responses, $L_i\in\{1,\ldots,L_{\max}\}$ denotes the number of responses processed for the $i$th intended virtual button, where $L_{\max}$ is the maximum allowed number of responses, and $b_i$ is the ground-truth label associated with response $x_i$. We define the average prediction set size $\overline{U}$ as the average number of plausible virtual-button labels included in the CP prediction sets:
$\overline{U}=\frac{1}{N_{\mathrm{cap}}}\sum_{x\in\mathcal{X}_{\mathrm{cap}}}\left|\mathcal{U}_{m}(x)\right|$.
A smaller $\overline{U}$ indicates that fewer button labels remain plausible, whereas a larger value indicates greater recognition uncertainty. Finally, the average number of re-inputs $\overline{R}$ is defined as
$\overline{R}=\frac{1}{N_{\mathrm{test}}}\sum_{i=1}^{N_{\mathrm{test}}}\left(L_i-1\right)$.

\subsection{Conformal Calibration and Performance Evaluation}
Next, we investigate the effect of the CP risk level $\alpha$ on the proposed adaptive recognition framework using both the DL and Random Forest models. Within each of the CC, CR, and CCR evaluation sets, $10\%$ of the labeled responses are reserved for CP calibration, $10\%$ are used as a validation subset for selecting $\alpha$, the acceptance threshold $\tau$, and the maximum capture budget $L_{\max}$, and $80\%$ are kept unseen for final evaluation. Using the validation subset, we vary $\alpha$ from $0.05$ to $0.30$ to examine its effect on the adaptive recognition framework. Figs.~\ref{fig:dl_adaptive_performance} and~\ref{fig:rf_adaptive_performance} illustrate the resulting performance trends for the DL and Random Forest models, respectively. As shown in Figs.~\ref{fig:empirical_coverage_dl} and~\ref{fig:prediction_set_size_dl}, increasing $\alpha$ reduces both empirical coverage and average prediction set size for the DL model. Lower risk levels produce larger, more conservative prediction sets, whereas higher risk levels produce smaller sets. CCR exhibits the lowest empirical coverage, consistent with the combined card and reader distribution shifts. The Random Forest model exhibits similar coverage and prediction set size trends, as shown in Figs.~\ref{fig:empirical_coverage_rf} and~\ref{fig:prediction_set_size_rf}. Adaptive recognition accuracy follows the same trend, as shown in Figs.~\ref{fig:adaptive_accuracy_dl} and~\ref{fig:adaptive_accuracy_rf}. Accuracy remains high at low and moderate risk levels but gradually decreases as $\alpha$ increases because smaller prediction sets are more likely to exclude the true button label. The DL model consistently outperforms Random Forest, while CCR yields the lowest accuracy for both models, consistent with the combined card and reader shift. Figs.~\ref{fig:average_reinputs_dl} and~\ref{fig:average_reinputs_rf} further show that the average number of re-inputs decreases substantially as $\alpha$ increases. Smaller prediction sets provide stronger evidence for included buttons, allowing the cumulative score to reach the acceptance threshold with fewer captures. The DL model generally requires fewer re-inputs because it produces more concentrated evidence. Overall, lower risk levels improve coverage and accuracy but require larger prediction sets and more re-inputs, whereas higher risk levels reduce interaction overhead at the cost of recognition reliability.  

\begin{table*}[!t]
\centering
\caption{Comparison of multi-capture decision methods.}
\label{tab:adaptive_baseline_comparison}

\small
\setlength{\tabcolsep}{3.5pt}
\renewcommand{\arraystretch}{1.2}

\begin{tabular}{@{}llccccc@{}}
\hline
\multirow{2}{*}{\textbf{Model}} &
\multirow{2}{*}{\textbf{Recognition Method}} &
\multicolumn{3}{c}{\textbf{Recognition Accuracy (\%)}} &
\multirow{2}{*}{\shortstack{\textbf{Adaptive}\\\textbf{Re-input}}} &
\multirow{2}{*}{\shortstack{\textbf{Avg.}\\\textbf{Re-inputs}}} \\
\cline{3-5}

& & \textbf{CC} & \textbf{CR} & \textbf{CCR} & & \\
\hline

\multirow{4}{*}{DL}
& Majority Voting
& $96.49 \pm 0.28$
& $96.50 \pm 0.31$
& $91.54 \pm 0.40$
& --
& \multirow{2}{*}{2.00} \\

& Probability Averaging
& $96.61 \pm 0.31$
& $96.62 \pm 0.31$
& $91.64 \pm 0.38$
& --
& \\

& Confidence-Based Adaptive
& $96.36 \pm 0.36$
& $96.33 \pm 0.39$
& $91.43 \pm 0.43$
& $\checkmark$
& 0.87 \\

& \textbf{CP-Based Adaptive (Ours)}
& $\boldsymbol{97.40 \pm 0.19}$
& $\boldsymbol{97.42 \pm 1.11}$
& $\boldsymbol{95.74 \pm 1.18}$
& $\checkmark$
& \textbf{0.55} \\

\hline

\multirow{4}{*}{Random Forest}
& Majority Voting
& $88.40 \pm 0.72$
& $89.90 \pm 0.85$
& $87.10 \pm 1.00$
& --
& \multirow{2}{*}{2.00} \\

& Probability Averaging
& $88.65 \pm 0.65$
& $90.35 \pm 0.92$
& $87.34 \pm 0.94$
& --
& \\

& Confidence-Based Adaptive
& $88.95 \pm 0.56$
& $91.00 \pm 1.60$
& $88.35 \pm 0.87$
& $\checkmark$
& 1.55 \\

& \textbf{CP-Based Adaptive (Ours)}
& $\boldsymbol{90.00 \pm 0.87}$
& $\boldsymbol{91.65 \pm 0.92}$
& $\boldsymbol{89.99 \pm 1.11}$
& $\checkmark$
& \textbf{1.03} \\

\hline
\end{tabular}
 % \vspace{-10pt}
\end{table*}

\subsubsection{Performance Across Domains}
To evaluate generalization, we use the $48{,}000$ previously unseen responses as the final held-out test data. The CC subset is drawn from Cards~16--25 collected using reader configuration $\mathcal{R}_1$, the CR subset is drawn from the source-domain Cards~1--5 collected using the previously unseen reader configuration $\mathcal{R}_3$, and the CCR subset is drawn from Cards~26--35 collected using reader configuration $\mathcal{R}_3$. None of these responses are used for domain-adaptation training, CP calibration, or operating-parameter selection. Using the validation subset, we evaluate the risk level over $\alpha=0.05$ to $0.30$, the acceptance threshold over $\tau\in [0.50,1.00]$, and the considered maximum capture budgets. The validation subset is used to select these operating parameters before evaluating the held-out test data. Based on the reliability--interaction tradeoff, we use $\alpha=0.07$ for DL and $\alpha=0.05$ for Random Forest, with $\tau=1.0$ and $L_{\max}=3$ for both. The selected value $L_{\max}=3$ limits user interaction while providing only marginal recognition gains beyond three captures. To isolate the effect of the CP-based adaptive recognition framework, we also evaluate the standard recognition accuracies of the domain-adapted DL and Random Forest models on the final held-out test data using only the initial response. As summarized in Table~\ref{tab:cp_adaptive_summary}, the DL and Random Forest models achieve average standard recognition accuracies of $93.15\%$ and $86.73\%$, respectively. With the CP-based adaptive recognition framework, the DL and Random Forest models achieve average empirical coverages of $97.71\%$ and $94.06\%$, respectively, and average adaptive recognition accuracies of $96.85\%$ and $90.55\%$. Their corresponding average prediction set sizes are $\overline{U}=1.34$ and $2.35$, while their average numbers of re-inputs are $\overline{R}=0.55$ and $1.03$, respectively. The DL model achieves higher empirical coverage and adaptive recognition accuracy while producing smaller prediction sets and requiring fewer re-inputs than the Random Forest model. The Random Forest results also show that the proposed CP-based adaptive recognition framework can be applied across different card--reader configurations and is not restricted to a specific model architecture.

\subsubsection{Computational Efficiency and Real-Time Feasibility} 
Next, we assess the computational efficiency of the proposed framework. Real-time inference is evaluated on a workstation equipped with a 24-core AMD Ryzen Threadripper 9960X CPU and an NVIDIA GeForce RTX 5090 GPU with 32~GB of memory. GNU Radio processing and CP-based adaptive decision-making are executed on the CPU, while DL inference is accelerated on the GPU. The DL model requires an average inference time of 18.71~ms per captured response, while the complete online computational pipeline requires an average of 65.57~ms to process each captured response and produce the corresponding accept-or-re-input decision. This measurement represents signal-processing, inference, and adaptive-decision latency after a response has been captured; it does not include the user-dependent time required to physically perform a re-input.

\begin{figure}[!t]
    \centering
    \includegraphics[width=\columnwidth]{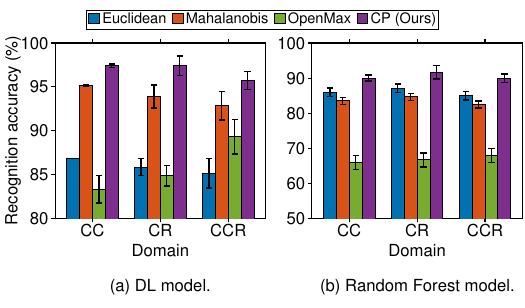}
    \caption{Recognition performance comparison using different methods.}
    \label{fig:methods_compare_combined}
    \vspace{-10pt}
\end{figure}

\subsection{Comparison with Different Recognition Methods} 
We compare four recognition methods using both the DL and Random Forest models: Euclidean distance-based recognition, Mahalanobis distance-based recognition, OpenMax, and the proposed CP-based adaptive method. Euclidean distance assigns each response to the nearest class centroid~\cite{10689531}, while Mahalanobis distance additionally accounts for the spread and correlation of the feature distributions~\cite{Venkataramanan2023GaussianLR,Sarpong2026CADRECD}. OpenMax uses Extreme Value Theory (EVT) by fitting a Weibull distribution to the tails of activation distances to identify responses that may not belong to the known classes~\cite{Bendale2015TowardsOS}. The proposed CP-based adaptive method instead constructs calibrated prediction sets and requests an additional capture only when the accumulated evidence is insufficient for reliable acceptance. As shown in Fig.~\ref{fig:methods_compare_combined}(a), the proposed CP-based adaptive method achieves the highest DL recognition accuracy across all three domains. For example, in the CC domain, CP achieves $97.40\%\pm0.19\%$, compared with $95.15\%\pm0.06\%$ for Mahalanobis distance and $83.32\%\pm1.56\%$ for OpenMax. In the more challenging CCR domain, CP maintains an accuracy of $95.74\%\pm1.18\%$, while Mahalanobis distance and Euclidean distance achieve $92.86\%\pm1.62\%$ and $85.14\%\pm1.67\%$, respectively. For the Random Forest model in Fig.~\ref{fig:methods_compare_combined}(b), the proposed CP-based adaptive method provides competitive recognition accuracy across the CC, CR, and CCR domains while enabling calibrated uncertainty estimation and adaptive re-input. These results demonstrate reliable adaptive recognition under card, reader, and combined domain variations.

\subsection{Comparison with Multi-Capture Decision Methods}
We further compare the proposed CP-based adaptive method with three conventional multi-capture decision methods: majority voting, probability averaging, and confidence-based adaptive recognition. All methods use the same pretrained DL or Random Forest model, with a maximum of three captures per button. Majority voting combines the predicted labels from three captures and selects the most frequently predicted button~\cite{Kuncheva2014}, while probability averaging combines their class probability scores and selects the button with the highest average probability~\cite{667881}. Both methods require two re-inputs for each button. Confidence-based adaptive recognition instead accepts a predicted button when its confidence exceeds a validation-selected threshold and requests additional captures otherwise, following the general principle of early classification~\cite{9564134}. In contrast, the proposed CP-based adaptive method uses calibrated prediction sets and accumulated evidence to determine whether an input can be accepted or requires repetition. As shown in Table~\ref{tab:adaptive_baseline_comparison}, the proposed method achieves the highest recognition accuracy across all three domains for both models while requiring fewer re-inputs. For the DL model, CP achieves $97.40\%\pm0.19\%$, $97.42\%\pm1.11\%$, and $95.74\%\pm1.18\%$ accuracy in the CC, CR, and CCR domains, respectively, compared with $96.36\%\pm0.36\%$, $96.33\%\pm0.39\%$, and $91.43\%\pm0.43\%$ for confidence-based adaptive recognition. Meanwhile, the average number of re-inputs decreases from 0.87 to 0.55. Similar improvements are observed for the Random Forest model, where CP improves recognition accuracy from $88.95\%$ to $90.00\%$ in CC, $91.00\%$ to $91.65\%$ in CR, and $88.35\%$ to $89.99\%$ in CCR, while reducing average re-inputs from 1.55 to 1.03. These results indicate that the proposed CP-based adaptive method provides a favorable balance between recognition accuracy and user interaction overhead across both learning models.

\subsection{Practical Deployment Considerations}
The proposed framework enables virtual PIN-button input on unmodified commercial NFC cards without altering ISO/IEC~15693 communication. However, practical deployment requires a reader capable of acquiring button-dependent RF signatures with sufficient sensitivity. The present SDR-based prototype uses separate transmit and receive paths, antenna tuning, and passive analog self-interference cancellation to acquire weak load-modulated responses. Integrating these capabilities into compact readers remains an important consideration. Recent work on impedance-based battery-free sensing~\cite{10.1145/3680207.3723480} and commodity-device sensing~\cite{295679} similarly highlights the importance of measurement robustness and hardware accessibility for practical RF-sensing systems.

Another consideration is complete PIN-entry reliability. Although the proposed framework evaluates individual-button recognition, practical PIN entry requires multiple consecutive correct decisions. Errors at individual positions can therefore affect the success of the complete PIN sequence. The adaptive recognition mechanism requests additional measurements when the available evidence is insufficient, reducing reliance on ambiguous single responses. Nevertheless, repeated interactions increase user effort, and computational processing latency alone does not represent complete PIN-entry time. Practical applications must therefore balance recognition reliability, interaction overhead, and user-dependent input duration. Prior battery-free sensing systems have also demonstrated the importance of maintaining reliable measurements under practical operating conditions~\cite{246304}.

Finally, changes in button-coil placement, orientation, nearby materials, and reader characteristics can perturb magnetic coupling and alter the measured response. Related work on magnetically controlled RFID interrogation~\cite{295619} and reader-dependent backscatter measurements~\cite{316056} illustrates the sensitivity of RF systems to their physical and acquisition conditions. The proposed domain adaptation framework addresses the evaluated card--reader variations, while conformal prediction supports adaptive decisions under the adopted calibration conditions. However, deployment in previously unseen environments may require additional calibration and validation. Future work will investigate compact-reader integration, uncontrolled user interactions, and complete-PIN performance to establish the operational requirements of the proposed interface.

\section{Conclusion}
\label{Conclude}
This paper presents a model-agnostic adaptive virtual PIN pad framework for passive NFC cards that integrates multi-domain adaptation and conformal prediction. By exploiting position-dependent magnetic coupling among the reader, NFC card, and passive button coil, the framework produces distinguishable RF responses for twelve virtual button regions without modifying commercial NFC ID cards or the ISO/IEC~15693 protocol. To address response variations across different cards and reader configurations, we employ a multi-domain adaptation framework to improve recognition robustness. Building on the adapted models, we develop a model-agnostic conformal prediction-based decision method that supports both deep learning and machine learning models and requests additional inputs when the accumulated evidence is insufficient for acceptance. Experimental results demonstrate improved cross-domain recognition accuracy and effective uncertainty-aware PIN-button recognition across different card--reader configurations.

% \newpage

% \section*{Acknowledgment}
% This work was supported in part by the National Science Foundation under grant no. CNS-2341846 and CNS-2310856. 

% \appendix
\bibliographystyle{IEEEtran}
% \small
\bibliography{ref}

@misc{iso15a, Key = {{ISO 15693}}, author = {{ISO, Geneva, Switzerland}}, type = {{Norm}}, title = {{Identification cards — Contactless integrated circuit cards — Vicinity cards}}, number = {{ISO 15693}}, year = {{2006}} }

@inproceedings{Ovadia2019TrustUncertainty,
  author    = {Yaniv Ovadia and Emily Fertig and Jie Ren and Zachary Nado and
               D. Sculley and Sebastian Nowozin and Joshua V. Dillon and
               Balaji Lakshminarayanan and Jasper Snoek},
  title     = {Can You Trust Your Model's Uncertainty? Evaluating Predictive Uncertainty Under Dataset Shift},
  booktitle = {Advances in Neural Information Processing Systems (NeurIPS)},
  volume    = {32},
  pages     = {13991--14002},
  year      = {2019}
}

@misc{CORDIER,
author = {Cordier, Thibault and Blot, Vincent and Lacombe, Louis and Morzadec, Thomas and Capitaine, Arnaud and Brunel, Nicolas},
year = {2023},
month = {09},
pages = {},
title = {Flexible and Systematic Uncertainty Estimation with Conformal Prediction via the MAPIE library}
}

@inproceedings {211253,
author = {Chuhan Gao and Yilong Li and Xinyu Zhang},
title = {{LiveTag}: Sensing {Human-Object} Interaction through Passive Chipless {WiFi} Tags},
booktitle = {15th USENIX Symposium on Networked Systems Design and Implementation (NSDI 18)},
year = {2018},
isbn = {978-1-939133-01-4},
address = {Renton, WA},
pages = {533--546},
publisher = {USENIX Association},
month = apr
}

@article{Haider2023AFP,
  title={A Fully Passive Ten-Digit Numeric Keypad Sensor Using Chipless RFID Technology},
  author={Usman A. Haider and Muhammad Noman and Hidayat Ullah and Hattan F. Abutarboush and Qammer Hussain Abbasi and Farooq Ahmad Tahir},
  journal={IEEE Sensors Journal},
  year={2023},
  volume={23},
  pages={2978-2987}
}

@InProceedings{pmlr-v235-huang24aa,
  title = 	 {Conformal Prediction for Deep Classifier via Label Ranking},
  author =       {Huang, Jianguo and Xi, Huajun and Zhang, Linjun and Yao, Huaxiu and Qiu, Yue and Wei, Hongxin},
  booktitle = 	 {Proceedings of the 41st International Conference on Machine Learning},
  pages = 	 {20331--20347},
  year = 	 {2024},
  editor = 	 {Salakhutdinov, Ruslan and Kolter, Zico and Heller, Katherine and Weller, Adrian and Oliver, Nuria and Scarlett, Jonathan and Berkenkamp, Felix},
  volume = 	 {235},
  series = 	 {Proceedings of Machine Learning Research},
  month = 	 {21--27 Jul},
  publisher =    {PMLR}
}

@InProceedings{pmlr-v235-kiyani24a,
  title = 	 {Conformal Prediction with Learned Features},
  author =       {Kiyani, Shayan and Pappas, George J. and Hassani, Hamed},
  booktitle = 	 {Proceedings of the 41st International Conference on Machine Learning},
  pages = 	 {24749--24769},
  year = 	 {2024},
  editor = 	 {Salakhutdinov, Ruslan and Kolter, Zico and Heller, Katherine and Weller, Adrian and Oliver, Nuria and Scarlett, Jonathan and Berkenkamp, Felix},
  volume = 	 {235},
  series = 	 {Proceedings of Machine Learning Research},
  month = 	 {21--27 Jul},
  publisher =    {PMLR}
}

@article{Chen2024CrossDeviceRF,
  title={Cross-Device Radio Frequency Fingerprinting Identification Based on Domain Adaptation},
  author={Zhiwei Chen and Zhibo Pang and Wenjing Hou and Hong Wen and Mi Wen and Runhui Zhao and Tao Tang},
  journal={IEEE Transactions on Consumer Electronics},
  year={2024},
  volume={70},
  pages={2391-2400}
}

@article{Yang2024MitigatingRI,
  title={Mitigating Receiver Impact on Radio Frequency Fingerprint Identification via Domain Adaptation},
  author={Liu Yang and Qiang Li and Xiaoyang Ren and Yi Fang and Shafei Wang},
  journal={IEEE Internet of Things Journal},
  year={2024},
  volume={11},
  pages={24024-24034}
}

@article{Shahid2020ChiplessRT,
  title={Chipless RFID Tag for Touch Event Sensing and Localization},
  author={Laiba Shahid and Humayun Shahid and Muhammad Ali Riaz and Syeda Iffat Naqvi and Muhammad Jamil Khan and Mansoor Shaukat Khan and Yasar Amin and Jonathan Loo},
  journal={IEEE Access},
  year={2020},
  volume={8},
  pages={502-513}
}

@misc{nxpicode2018,
  author       = {{NXP Semiconductors}},
  title        = {{ICODE SLIX – SL2S2002, ISO/IEC 15693 compliant smart label IC}},
  howpublished = {{Product Data Sheet, Rev. 3.7}},
  year         = {2018},
  month        = {Dec}
}

@article{Sarpong2026CADRECD,
  title={CADRE: Card-Agnostic Domain-Aligned RF Embeddings for Virtual PIN Pads on Passive NFC Cards},
  author={Dickson Akuoko Sarpong and Hongzhi Guo},
  journal={2026 22nd Annual IEEE International Conference on Sensing, Communication, and Networking (SECON)},
  year={2026},
  pages={332-340}
}

@article{10.1561/2200000101,
journal = {Foundations and Trends in Machine Learning},
author = {Angelopoulos, Anastasios N. and Bates, Stephen},
title = {Conformal Prediction: A Gentle Introduction},
year = {2023},
issue_date = {Mar 2023},
publisher = {Now Publishers Inc.},
address = {Hanover, MA, USA},
volume = {16},
number = {4},
issn = {1935-8237},
doi = {10.1561/2200000101},
month = mar,
pages = {494–591},
numpages = {114}
}

@inproceedings{10.1145/2702123.2702178,
author = {Li, Hanchuan and Ye, Can and Sample, Alanson P.},
title = {IDSense: A Human Object Interaction Detection System Based on Passive UHF RFID},
year = {2015},
isbn = {9781450331456},
publisher = {Association for Computing Machinery},
address = {New York, NY, USA},
booktitle = {Proceedings of the 33rd Annual ACM Conference on Human Factors in Computing Systems},
pages = {2555–2564},
numpages = {10},
location = {Seoul, Republic of Korea},
series = {CHI '15}
}

@inproceedings{10.1145/2858036.2858249,
author = {Li, Hanchuan and Brockmeyer, Eric and Carter, Elizabeth J. and Fromm, Josh and Hudson, Scott E. and Patel, Shwetak N. and Sample, Alanson},
title = {PaperID: A Technique for Drawing Functional Battery-Free Wireless Interfaces on Paper},
year = {2016},
isbn = {9781450333627},
publisher = {Association for Computing Machinery},
address = {New York, NY, USA},
booktitle = {Proceedings of the 2016 CHI Conference on Human Factors in Computing Systems},
pages = {5885–5896},
numpages = {12},
location = {San Jose, California, USA},
series = {CHI '16}
}

@inproceedings{10.1145/3332165.3347910,
author = {Hsieh, Meng-Ju and Guo, Jr-Ling and Lu, Chin-Yuan and Hsieh, Han-Wei and Liang, Rong-Hao and Chen, Bing-Yu},
title = {RFTouchPads: Batteryless and Wireless Modular Touch Sensor Pads Based on RFID},
year = {2019},
isbn = {9781450368162},
publisher = {Association for Computing Machinery},
address = {New York, NY, USA},
booktitle = {Proceedings of the 32nd Annual ACM Symposium on User Interface Software and Technology},
pages = {999–1011},
numpages = {13},
location = {New Orleans, LA, USA},
series = {UIST '19}
}

@INPROCEEDINGS{9155259,
  author={Al-Shawabka, Amani and Restuccia, Francesco and D’Oro, Salvatore and Jian, Tong and Costa Rendon, Bruno and Soltani, Nasim and Dy, Jennifer and Ioannidis, Stratis and Chowdhury, Kaushik and Melodia, Tommaso},
  booktitle={IEEE INFOCOM 2020 - IEEE Conference on Computer Communications}, 
  title={Exposing the Fingerprint: Dissecting the Impact of the Wireless Channel on Radio Fingerprinting}, 
  year={2020},
  volume={},
  number={},
  pages={646-655},
 doi={10.1109/INFOCOM41043.2020.9155259}}

@ARTICLE{10034841,
  author={Zha, Xiong and Li, Tianyun and Qiu, Zhaoyang and Li, Fang},
  journal={IEEE Signal Processing Letters}, 
  title={Cross-Receiver Radio Frequency Fingerprint Identification Based on Contrastive Learning and Subdomain Adaptation}, 
  year={2023},
  volume={30},
  number={},
  pages={70-74},
  doi={10.1109/LSP.2023.3241592}}

@ARTICLE{10345732,
  author={Shen, Guanxiong and Zhang, Junqing and Marshall, Alan and Woods, Roger and Cavallaro, Joseph and Chen, Liquan},
  journal={IEEE Transactions on Mobile Computing}, 
  title={Towards Receiver-Agnostic and Collaborative Radio Frequency Fingerprint Identification}, 
  year={2024},
  volume={23},
  number={7},
  pages={7618-7634},
  doi={10.1109/TMC.2023.3340039}}

@ARTICLE{10923619,
  author={Jiao, Wanguo and Du, Wei and Zhang, Changsheng and Suo, Long},
  journal={IEEE Sensors Journal}, 
  title={A High-Efficient Wi-Fi-Based Cross-Domain Recognition Framework Using Multisource Domain Adaptation for Single-Transceiver Scenarios}, 
  year={2025},
  volume={25},
  number={8},
  pages={14196-14208},
  doi={10.1109/JSEN.2025.3540664}}

@inproceedings{3495724.3496026,
booktitle = {Advances in Neural Information Processing Systems},
author = {Romano, Yaniv and Sesia, Matteo and Cand\`{e}s, Emmanuel J.},
title = {Classification with valid and adaptive coverage},
year = {2020},
isbn = {9781713829546},
publisher = {Curran Associates Inc.},
address = {Red Hook, NY, USA},
articleno = {302},
numpages = {11},
location = {Vancouver, BC, Canada},
series = {NIPS '20}
}

@inproceedings{10.1145/3714394.3754366,
author = {Ma, Chongxuan and Xie, Zheng and Guo, Kaiwen and Liu, Chao},
title = {NFINGER: Personalized Backside Gesture Recognition on Smartphones Using Back-mounted NFC Tag Array},
year = {2026},
isbn = {9798400714771},
publisher = {Association for Computing Machinery},
address = {New York, NY, USA},
doi = {10.1145/3714394.3754366},
booktitle = {Companion of the 2025 ACM International Joint Conference on Pervasive and Ubiquitous Computing},
pages = {176–180},
numpages = {5},
location = {Finland},
series = {UbiComp Companion '25}
}

@INPROCEEDINGS{8737419,
  author={Zhao, Cui and Li, Zhenjiang and Liu, Ting and Ding, Han and Han, Jinsong and Xi, Wei and Gui, Ruowei},
  booktitle={IEEE INFOCOM 2019 - IEEE Conference on Computer Communications}, 
  title={RF-Mehndi: A Fingertip Profiled RF Identifier}, 
  year={2019},
  volume={},
  number={},
  pages={1513-1521},
  doi={10.1109/INFOCOM.2019.8737419}}

@article{Bendale2015TowardsOS,
  title={Towards Open Set Deep Networks},
  author={Abhijit Bendale and Terrance E. Boult},
  journal={2016 IEEE Conference on Computer Vision and Pattern Recognition (CVPR)},
  year={2016},
  pages={1563-1572}
}

@inproceedings{Sun2016DeepCC,
  title={Deep CORAL: Correlation Alignment for Deep Domain Adaptation},
  author={Baochen Sun and Kate Saenko},
  booktitle={ECCV Workshops},
  year={2016}
}

@ARTICLE{8434290,
  author={Zhang, Yun and Wang, Nianbin and Cai, Shaobin and Song, Lei},
  journal={IEEE Access}, 
  title={Unsupervised Domain Adaptation by Mapped Correlation Alignment}, 
  year={2018},
  volume={6},
  number={},
  pages={44698-44706},
  doi={10.1109/ACCESS.2018.2865249}}

@article{Sarpong2025ModelAgnosticUQ,
  title={Model-Agnostic Uncertainty Quantification for Fast NFC Tag Identification Using RF Fingerprinting},
  author={Dickson Akuoko Sarpong and Adam Kamrath and Rohit Bhusal and Hongzhi Guo},
  journal={IEEE Internet of Things Journal},
  year={2025},
  volume={12},
  pages={47607-47622}
}

@article{de2021adapt,
	  title={ADAPT: Awesome Domain Adaptation Python Toolbox},
	  author={de Mathelin, Antoine and Deheeger, Fran{\c{c}}ois and Richard, Guillaume and Mougeot, Mathilde and Vayatis, Nicolas},
	  journal={arXiv preprint arXiv:2107.03049},
	  year={2021}
	}

@article{Sun2016CorrelationAF,
  title={Correlation Alignment for Unsupervised Domain Adaptation},
  author={Baochen Sun and Jiashi Feng and Kate Saenko},
  journal={ArXiv},
  year={2016},
  volume={abs/1612.01939}
}

@INPROCEEDINGS{11044651,
  author={Xie, Lingnan and Peng, Linning and Zhang, Junqing},
  booktitle={IEEE INFOCOM 2025 - IEEE Conference on Computer Communications}, 
  title={Towards Robust RF Fingerprint Identification Using Spectral Regrowth and Carrier Frequency Offset}, 
  year={2025},
  volume={},
  number={},
  pages={1-10},
  doi={10.1109/INFOCOM55648.2025.11044651}}

@INPROCEEDINGS{11044470,
  author={Tong, Jingyu and Zhao, Xiaopeng and Wang, Zhicheng and Dai, Donghui and An, Zhenlin and Yang, Lei},
  booktitle={IEEE INFOCOM 2025 - IEEE Conference on Computer Communications}, 
  title={Commercial RFIDs as Reconfigurable Intelligent Surfaces}, 
  year={2025},
  volume={},
  number={},
  pages={1-10},
  doi={10.1109/INFOCOM55648.2025.11044470}}

@INPROCEEDINGS{9155251,
  author={Yang, Panlong and Feng, Yuanhao and Xiong, Jie and Chen, Ziyang and Li, Xiang-Yang},
  booktitle={IEEE INFOCOM 2020 - IEEE Conference on Computer Communications}, 
  title={RF-Ear: Contactless Multi-device Vibration Sensing and Identification Using COTS RFID}, 
  year={2020},
  volume={},
  number={},
  pages={297-306},
  doi={10.1109/INFOCOM41043.2020.9155251}}

@inproceedings{10.1145/3117811.3117818,
author = {Pradhan, Swadhin and Chai, Eugene and Sundaresan, Karthikeyan and Qiu, Lili and Khojastepour, Mohammad A. and Rangarajan, Sampath},
title = {RIO: A Pervasive RFID-based Touch Gesture Interface},
year = {2017},
isbn = {9781450349161},
publisher = {Association for Computing Machinery},
address = {New York, NY, USA},
booktitle = {Proceedings of the 23rd Annual International Conference on Mobile Computing and Networking},
pages = {261–274},
numpages = {14},
location = {Snowbird, Utah, USA},
series = {MobiCom '17}
}

@article{10.1145/3631442,
  author    = {John Nolan and Kun Qian and Xinyu Zhang},
  title     = {KeyStub: A Passive RFID-based Keypad Interface Using Resonant Stubs},
  journal   = {ACM Transactions on Computer-Human Interaction},
  year      = {2024},
  volume    = {31},
  number    = {1},
  articleno = {173},
  numpages  = {23},
  publisher = {Association for Computing Machinery},
  address   = {New York, NY, USA},
  doi       = {10.1145/3631442},
  month     = jan
}

@ARTICLE{10689531,
  author={Zhang, Lingfeng and Wan, Zunian and Ding, Yepeng and Hu, Tao and Ogawa, Takefumi and Sato, Hiroyuki},
  journal={IEEE Access}, 
  title={Hand Gesture Classification Using sEMG Signals: Nearest-Centroid-Based Methodology With DBA}, 
  year={2024},
  volume={12},
  number={},
  pages={141916-141931},
  doi={10.1109/ACCESS.2024.3467158}}

@article{Venkataramanan2023GaussianLR,
  title={Gaussian Latent Representations for Uncertainty Estimation using Mahalanobis Distance in Deep Classifiers},
  author={Aishwarya Venkataramanan and Assia Benbihi and Martin Laviale and C{\'e}dric Pradalier},
  journal={2023 IEEE/CVF International Conference on Computer Vision Workshops (ICCVW)},
  year={2023},
  pages={4490-4499}
}

@INPROCEEDINGS{8486346,
  author={Wang, Chuyu and Liu, Jian and Chen, Yingying and Liu, Hongbo and Xie, Lei and Wang, Wei and He, Bingbing and Lu, Sanglu},
  booktitle={IEEE INFOCOM 2018 - IEEE Conference on Computer Communications}, 
  title={Multi - Touch in the Air: Device-Free Finger Tracking and Gesture Recognition via COTS RFID}, 
  year={2018},
  volume={},
  number={},
  pages={1691-1699},
  doi={10.1109/INFOCOM.2018.8486346}}

@inproceedings{10.1145/3680207.3723480,
author = {Li, Liyao and Wu, Yun and Kim, Minsung and Shang, Bozhao and Xiong, Jie and Xu, Wenyao and Chen, Xiaojiang and Xie, Yaxiong},
title = {From Signal-based to Impedance-based Sensing: A paradigm Shift for Plug-and-Play, Mobile, and Sensitive Battery-free Sensing},
year = {2025},
isbn = {9798400711299},
publisher = {Association for Computing Machinery},
address = {New York, NY, USA},
doi = {10.1145/3680207.3723480},
booktitle = {Proceedings of the 31st Annual International Conference on Mobile Computing and Networking},
pages = {375–390},
numpages = {16},
location = {Hong Kong, China},
series = {ACM MOBICOM '25}
}

@inproceedings {295679,
author = {Enze Yi and Dan Wu and Jie Xiong and Fusang Zhang and Kai Niu and Wenwei Li and Daqing Zhang},
title = {{BFMSense}: {WiFi} Sensing Using Beamforming Feedback Matrix},
booktitle = {21st USENIX Symposium on Networked Systems Design and Implementation (NSDI 24)},
year = {2024},
isbn = {978-1-939133-39-7},
address = {Santa Clara, CA},
pages = {1697--1712},
publisher = {USENIX Association},
month = apr
}

@inproceedings {295619,
author = {Donghui Dai and Zhenlin An and Zheng Gong and Qingrui Pan and Lei Yang},
title = {{RFID+}: Spatially Controllable Identification of {UHF} {RFIDs} via Controlled Magnetic Fields},
booktitle = {21st USENIX Symposium on Networked Systems Design and Implementation (NSDI 24)},
year = {2024},
isbn = {978-1-939133-39-7},
address = {Santa Clara, CA},
pages = {1351--1367},
publisher = {USENIX Association},
month = apr
}

@inproceedings {316056,
author = {Jia Liu and Yifei Ma and Xingyu Chen and Haipeng Dai and He Huang and Zihao Lin and Wei Zheng and Junzhao Du and Guihai Chen},
title = {Decoding {RSSI} Compression in {RFID}: Dynamic {RCS} Modeling and {Tag-Intrinsic} Power Metrics for Reliable Backscatter Networks},
booktitle = {23rd USENIX Symposium on Networked Systems Design and Implementation (NSDI 26)},
year = {2026},
isbn = {978-1-939133-54-0},
address = {Renton, WA},
pages = {253--266},
publisher = {USENIX Association},
month = may
}

@inproceedings {246304,
author = {Unsoo Ha and Junshan Leng and Alaa Khaddaj and Fadel Adib},
title = {Food and Liquid Sensing in Practical Environments using {RFIDs} },
booktitle = {17th USENIX Symposium on Networked Systems Design and Implementation (NSDI 20)},
year = {2020},
isbn = {978-1-939133-13-7},
address = {Santa Clara, CA},
pages = {1083--1100},
publisher = {USENIX Association},
month = feb
}

@inbook{Kuncheva2014,
  author    = {Kuncheva, Ludmila I.},
  title     = {Combining Label Outputs},
  booktitle = {Combining Pattern Classifiers: Methods and Algorithms},
  edition   = {2},
  chapter   = {4},
  pages     = {111--142},
  publisher = {John Wiley \& Sons},
  year      = {2014},
  doi       = {10.1002/9781118914564.ch4}
}

@techreport{2022PersonalIV,
  author      = {{National Institute of Standards and Technology}},
  title       = {Personal Identity Verification ({PIV}) of Federal Employees and Contractors},
  institution = {National Institute of Standards and Technology},
  type        = {Federal Information Processing Standards Publication},
  number      = {FIPS 201-3},
  year        = {2022},
  month       = jan,
  doi         = {10.6028/NIST.FIPS.201-3}
}

@misc{VirtualPINPad2026,
  title = {Source code and implementation resources},
  year  = {2026},
  url   = {https://github.com/dicksonakuoko/Virtual-NFC-Card-PIN-Pad}
}

@ARTICLE{667881,
  author={Kittler, J. and Hatef, M. and Duin, R.P.W. and Matas, J.},
  journal={IEEE Transactions on Pattern Analysis and Machine Intelligence}, 
  title={On combining classifiers}, 
  year={1998},
  volume={20},
  number={3},
  pages={226-239},
  doi={10.1109/34.667881}}

@INPROCEEDINGS{9564134,
  author={Zafar, Paul-Emile and Achenchabe, Youssef and Bondu, Alexis and Cornuéjols, Antoine and Lemaire, Vincent},
  booktitle={2021 IEEE 8th International Conference on Data Science and Advanced Analytics (DSAA)}, 
  title={Early Classification of Time Series: Cost-based multiclass Algorithms}, 
  year={2021},
  volume={},
  number={},
  pages={1-10},
  doi={10.1109/DSAA53316.2021.9564134}}

\end{document}